\documentclass[journal]{IEEEtran}
\renewcommand\IEEEkeywordsname{Index Terms}
\usepackage{blindtext}
\usepackage{graphicx}
\usepackage{amssymb}
\usepackage{amsmath}
\usepackage{amsfonts}
\usepackage{xcolor}
\usepackage{braket}
\usepackage{multirow}
\usepackage{subfigure}
\usepackage{cite}
\usepackage[ruled]{algorithm2e}
\usepackage[utf8]{inputenc}
\usepackage[english]{babel}
\usepackage{caption}
\usepackage{amsthm}
\ifCLASSINFOpdf
\else
\fi
\newcommand{\vh}{{\bf h}}

\newcommand{\vv}{{\bf v}}

\newcommand{\cR}{{\cal R}}
\newcommand{\cS}{{\cal S}}
\newcommand{\cT}{{\cal T}}

\newcommand{\bN}{{\mathbb{N}}}

\newenvironment{mat}[1]{\left[\begin{array}{#1}}{\end{array}\right]}

\renewcommand{\tt}[2]{{\cT_{#1}^{#2}}}

\newcommand{\ignore}[1]{}

\begin{document}
%
\newtheorem{corol}[theorem]{Corollary}
\title{A Review on Coexistence Issues of Broadband Millimeter-Wave Communications}
%
%
%

\author{
Ching-Lun Tai$^1$ and~Derek Wu$^1$\\
\quad\\
$^1$School of Electrical and Computer Engineering, Georgia Institute of Technology, GA, United States
}

\twocolumn[
\begin{@twocolumnfalse}
\maketitle

\begin{abstract}
With higher frequencies and broader spectrum than conventional frequency bands, the millimeter-wave (mmWave) band is suitable for next-generation wireless networks featuring short-distance high-rate communications.
As a newcomer, mmWaves are expected to have the backward compatibility with existing services and collaborate with other technologies in order to enhance system performances.
Therefore, the coexistence issues become an essential topic for next-generation wireless communications.
In this paper, we systematically review the coexistence issues of broadband mmWave communications and their corresponding solutions proposed in the literature, helping shed light on the insights of the mmWave design.
Particularly, the works surveyed in this paper can be classified into four categories: coexistence with microwave communications, coexistence with fixed services, coexistence with non-orthogonal multiple access (NOMA), and other coexistence issues.
Results of numerical evaluations inspired by the literature are presented for a deeper analysis.
We also point out some challenges and future directions for each category as a roadmap to further investigate the coexistence issues of broadband mmWave communications.

\end{abstract}

\begin{IEEEkeywords}
Millimeter-wave (mmWave), coexistence, microwave, fixed services, non-orthogonal multiple access (NOMA)
\end{IEEEkeywords}
\vspace{0.5cm}
\end{@twocolumnfalse}
]

%
\IEEEpeerreviewmaketitle

\section{Introduction}
With a growing demand of wireless connected devices, such as tablets, smartphones, and internet of things (IoT) equipment \cite{xia12}, the industry of wireless communications prospers rapidly in recent years \cite{rangan14}. 
However, the radio frequency (RF) resources are quite limited over the conventional congested frequency band \cite{pi11}.
Therefore, it is natural to look into higher frequencies for a broader available spectrum, which has not yet been fully licensed or occupied by various wireless services.
Particularly, the frequency band of 24.25-86 GHz, which belongs to the millimeter-wave (mmWave) spectrum, is considered as a promising solution \cite{administrations15}, featuring a sufficient bandwidth which leads to a significantly increased rate and capacity for wireless communications.
In general, we refer to the communications over the 24.25-86 GHz band as mmWave communications.

Compared with lower-frequency waves, mmWaves have several unique characteristics.
Due to their short wavelengths, mmWaves suffer from a blockage effect, which prevents them from penetrating thicker glass or walls \cite{rangan14}.
Moreover, some of the mmWave bands are susceptible to a severe rain or oxygen attenuation \cite{li06,wells09,rappaport13}.
Consequently, mmWaves are subject to a larger path loss than lower-frequency waves during propagation, and hence mmWaves are more suitable for short-distance communications.

As the number of devices and the demand of high data rates increase, the ultra-dense networks \cite{kamel16} with short-distance high-rate transmissions are expected to be deployed in next-generation wireless communications, where mmWaves are regarded as a key technology.

Despite the advantages of mmWave communications, not all existing systems are compatible with mmWaves.
Besides, over the mmWave band are some existing fixed services, and the incoming cellular networks with mmWave communications and these fixed services will interfere with each other.
Moreover, new proposed technologies can be combined with mmWave communications in the network design.
Therefore, the coexistence issues of mmWave communications are essential to next-generation wireless communications, not only because of the backward compatibility required during the transition, but also because of the promising improvement of system performances, and these issues are the main focus in this paper.



The contribution of this paper is to provide a systematic review of the coexistence issues of broadband mmWave communications and their corresponding solutions proposed in the literature.
To the best of our knowledge, this is the first work that systematically introduces the recent advancements in these issues and helps shed light on the insights of mmWave system design.


\begin{figure*}[ht]
\centering
\includegraphics[width=15cm]{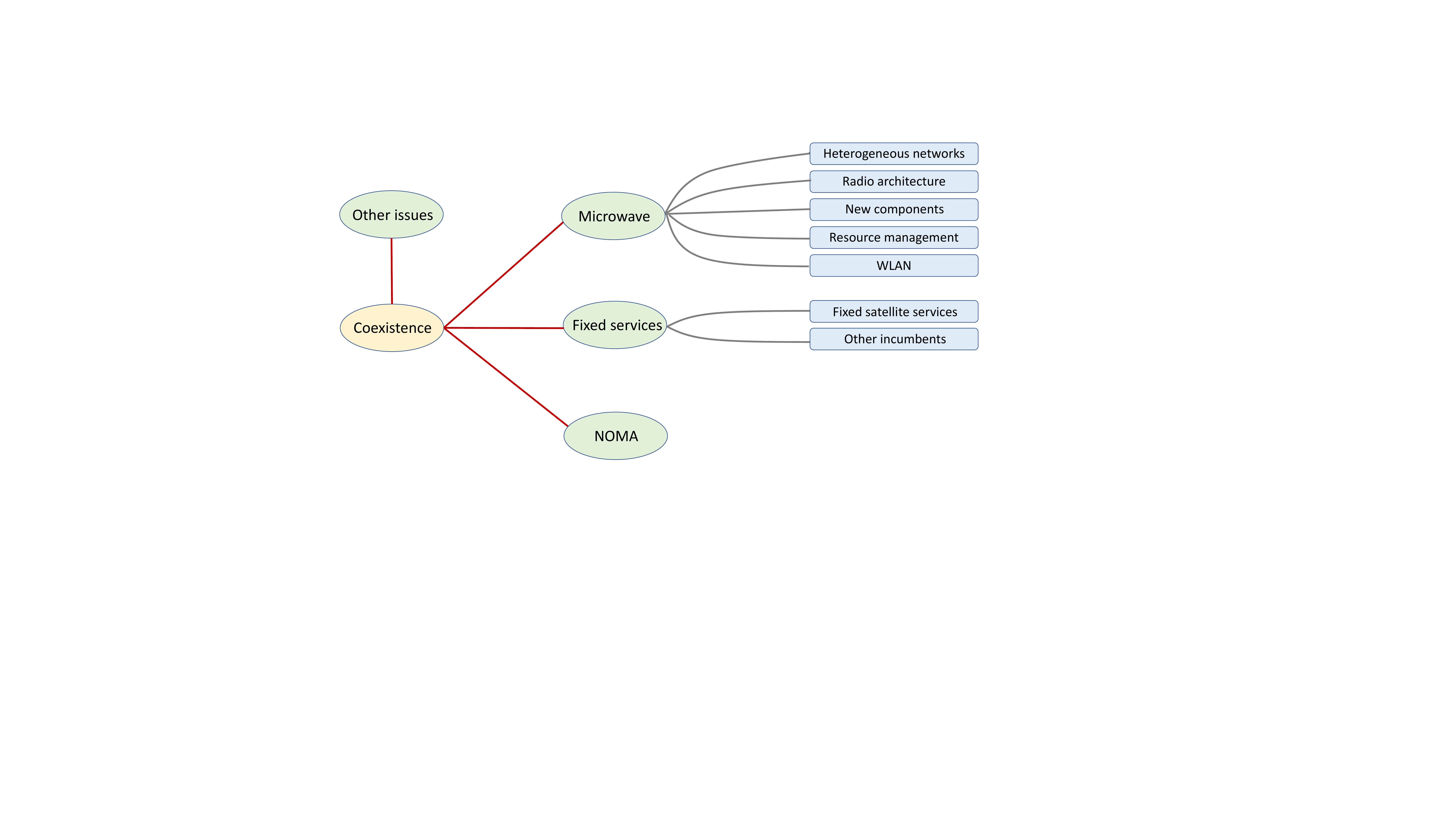}
\caption{Coexistence issues of broadband mmWave communications}
\label{fig:summary}
\end{figure*}

Particularly, the coexistence issues reviewed in this paper can be classified into the following four categories (which are summarized in Fig. \ref{fig:summary}):


\begin{itemize}
    \item{\textbf{Coexistence with microwave communications}:
    Conventional wireless networks adopt microwave communications, which generally work in the frequency band ranging from 0.3 GHz to 6 GHz (as known as sub-6 GHz) and suffer from the limited bandwidth.
    Due to their non-overlapping frequency bands, the radio interference is not a significant problem for the coexistence of mmWave and microwave communications.
    Instead, the mutual cooperation and novel design of both communications to enhance the system performances are the main focuses in this category.
    
    }
    \item{\textbf{Coexistence with fixed services}: 
    Over the mmWave band, there are various existing fixed services, including fixed satellite services and other incumbents.
    However, the incoming cellular networks with mmWave communications and these fixed services will interfere with each other.
    In this situation, there are an interferer (cellular networks or fixed services) and a victim (fixed services or cellular networks).
    Therefore, the interference mitigation mechanisms which reduce the interference to an acceptable level are the main focus in this category.
    
    }
    \item{\textbf{Coexistence with non-orthogonal multiple access (NOMA)}: 
    NOMA is a technique which multiplexes the signals in the power domain, featuring enhanced spectrum sharing and power allocation \cite{saito13,dai15,liu17}.
    Due to its characteristics, NOMA can be combined with mmWave communications to improve the multi-connectivity of wireless networks \cite{ding17}.
    Therefore, the combination of mmWaves and NOMA is the main focus in this category.
    }
    \item{\textbf{Other coexistence issues}:
    In addition to the research in the above categories, there are several works focusing on other discrete topics of the coexistence issues of broadband mmWave communications from different perspectives.
    The main focus in this category is to present the diversified research directions toward the coexistence issues.
    
    }
\end{itemize}

The remainder of this paper is organized as follows.
We review the coexistence of mmWave and microwave communications in Sec. \ref{sec:microwave}. 
In Sec. \ref{sec:FS}, the coexistence of mmWave communications and fixed services is studied. 
The coexistence of mmWave communications and NOMA is investigated in Sec. \ref{sec:NOMA}. 
In Sec. \ref{sec:other}, other coexistence issues are introduced. 
Results of numerical evaluations are presented in Sec. \ref{sec:simu}. 
In Sec. \ref{sec:challenges}, we point out the challenges and future directions of the coexistence issues.
Finally, Sec. \ref{sec:conclusion} concludes the paper.


\section{Coexistence with Microwave Communications}
\label{sec:microwave}
\begin{figure*}[ht]
\centering
\includegraphics[width=16cm]{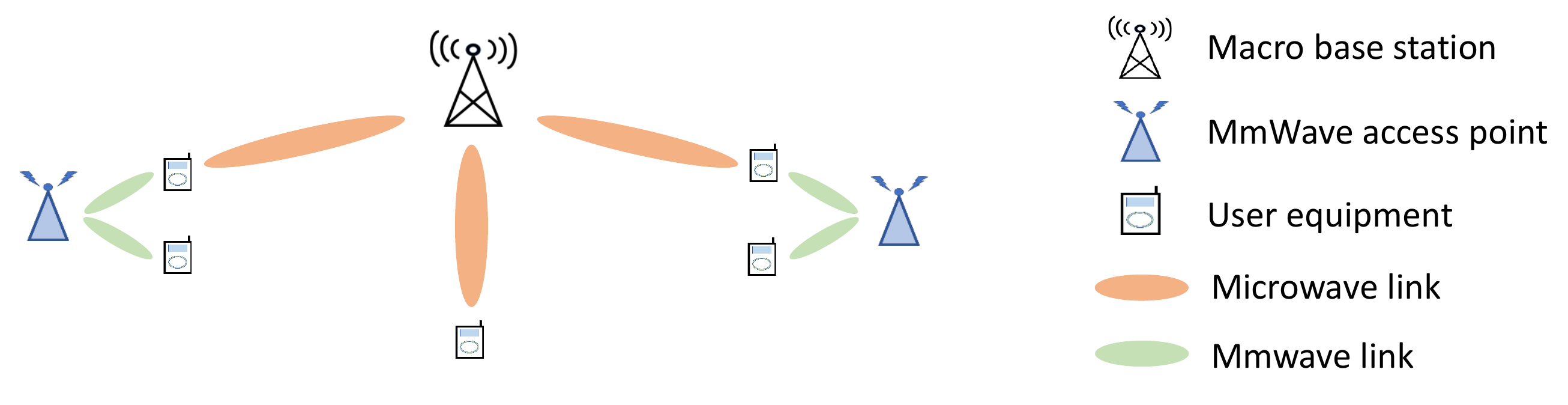}
\caption{Illustration of coexistence of mmWave and microwave communications}
\label{fig:cat1}
\end{figure*}

In this section, we review the related issues about the coexistence of mmWave and microwave communications.
An illustration of coexistence of both communications is shown in Fig. \ref{fig:cat1}, where user equipment can be supported by mmWave and microwave links through the mmWave access points and macro base station.




\subsection{Coexistence in Heterogeneous Networks}
In heterogeneous networks, mmWave and microwave communications can not only coexist but also cooperate for better networking performances.
Considering the different characteristics of these two kinds of communications, several aspects need to be taken into account in order to put them into right roles.

There are three critical perspectives when investigating heterogeneous networks: physical layer security, content caching, and wireless energy harvesting \cite{wang18}.

First, physical layer security is developed as a low-complexity measure for secure communications when malicious eavesdroppers are present \cite{zheng13,mukherjee14,yang15b,zou15,zhu17}.
With the base station densification, the physical layer security of heterogeneous networks can be improved, since legitimate users will benefit from enhanced spectral efficiency \cite{andrews14}, while eavesdroppers relying on side channels will suffer from increased interference.

Second, content caching is required in next-generation communications, where content caches should be placed near to user equipment \cite{wang18}.
Therefore, the development of content-aware user association is of great importance in order to reduce backhaul latency, and the link designation for content delivery needs to be determined based on the characteristics of both mmWave and microwave communications \cite{wang18}.

In \cite{zhu18}, the authors investigate the content placement in heterogeneous networks with multi-antenna base stations.
Based on the probabilistic caching model in \cite{blaszczyszyn15} and the average line-of-sight (LoS) model in \cite{bai15,park16}, the successful content delivery probability (SCDP) of multi-antenna base stations in both mmWave and microwave communications is derived, depending on channel effects, caching placement probability, base station density, transmit power, and antenna number \cite{zhu18}.
In addition, \cite{zhu18} proposes a constrained cross-entropy optimization algorithm and a heuristic two-stage algorithm in order to maximize the SCDP.

Third, wireless energy harvesting is required to prolong the lifetime of user equipment \cite{wang18}.
Note that the efficiency of wireless energy harvesting depends on both RF-to-DC conversion and path loss.
Therefore, rectifier design for high RF-to-DC conversion efficiency and adequate distance between base stations and user equipment for tolerable path loss are crucial \cite{wang18}.

In \cite{wang17}, the wireless energy harvesting in heterogeneous networks is investigated.
The energy coverage probability regarding directed transferred power and ambient RF harvested power in both mmWave and microwave tiers is analyzed in \cite{wang17}.
It is shown that the power transfer connectivity probability depends on the density of base stations \cite{wang17}.

In next-generation heterogeneous networks, one of the most important applications is the intelligent transport system (ITS), which is aimed at enabling safer and more efficient road experiences.
ITS mainly involves the design of vehicular networks, including both vehicle-to-vehicle (V2V) and vehicle-to-everything (V2X) communications which can be implemented through the technologies of both wireless communications and IoT \cite{An11}.

In \cite{wu17}, the authors investigate the content delivery in heterogeneous vehicular networks, where mmWave links are used for V2V communication, and microwave links are used for both V2V communication and connection between base stations and vehicles.
In addition, \cite{wu17} proposes an efficient fuzzy logic \cite{zadeh88} based algorithm selecting a predetermined number of vehicles as gateway heads, which communicate with both the base station and neighboring vehicles within a specific range, according to the competency value, which depends on three factors: velocity, moving direction, and antenna height.

In \cite{anjinappa18}, the angular and temporal correlation between mmWave and microwave bands for V2X channels is investigated.
Suppose there are $K(t)$ paths in a V2X channel, where $t$ is the time when the channel is excited.
Then, the omni-directional channel impulse response (CIR) given frequency $f$ can be expressed as
{\small
\begin{equation}
    h(t,\tau,\psi,\theta;f)=\sum_{k=1}^{K(t)}\rho_k e^{-j\phi_k}\delta(\tau-\tau_k)\delta(\psi-\psi_k)\delta(\theta-\theta_k),
\end{equation}
}where $\tau$ is the time delay, $\psi$ is the angle of arrival, $\theta$ is the angle of departure, and $\rho_k$, $\phi_k$, $\tau_k$, $\psi_k$, and $\theta_k$ are the amplitude, phase, time delay, angle of arrival, and angle of departure associated with the $k$th path, respectively.
If the CIR given one frequency is strong correlated to the CIR given another frequency, then there is a strong angular and temporal correlation between these two frequencies.
Particularly, it is found that there is a strong angular and temporal correlation between 28 GHz in the mmWave band and 5.9 GHz in the microwave band \cite{anjinappa18}.

Recently, different scenarios of heterogeneous networks have been studied in the literature.

In \cite{rois16}, a heterogeneous network featuring full-dimension massive input massive output (FD-MIMO) \cite{kim14b}, where 2D planar antenna arrays enable 3D beamforming to specific user equipment at the base station, is proposed.
In this network, mmWave systems adopt beam-steerable narrow-beam antennas \cite{rappaport13} and a sufficient number of access points, which serve as relay nodes (RNs), to allow multi-hop communication and LoS connectivity \cite{rois15}, while microwave systems adopt carrier aggregation (CA) \cite{pedersen11} to increase the data rate.
In order to enhance the benefits of this network, the authors of \cite{rois16} adopt the service-driven dynamic radio resource management for proper resource allocation and propose a protocol called multi-layered dynamic transmission scheme, where at any specific time slot, the network will be dynamically sliced into different layers, with each layer involving part of network nodes and operating with either mmWave or microwave links.

In \cite{deng17}, a heterogeneous network where mmWave and microwave carriers coexist in multi-connectivity phantom cells \cite{ishii12} is considered.
The authors of \cite{deng17} propose a hierarchical architecture, where the user equipment serves as relays to enable two-hop relaying \cite{osseiran14} and which consists of a logical central coordinator, local base station controllers, and a cooperative device-to-device (D2D) network, for beam discovery, channel measurement, relay selection, resource allocation, and interference coordination.
For beam recovery and channel measurement, \cite{deng17} adopts the transmission/reception silencing patterns in \cite{tiirola13} to enhance the speed and scalability.
For relay selection, resource allocation, and interference coordination, \cite{deng17} adopts interference graphs to model interactions between neighboring links and manages them with the concept of graph coloring \cite{chaitin82}. 

In \cite{busari18}, a heterogeneous network where mmWave links are used for small cells with bandwidth-hungry applications \cite{busari17} and microwave links are used for macrocells, is evaluated under LoS, non-LoS (NLoS), and outdoor-to-indoor (O2I) propagation environments, respectively, in a multi-user downlink scenario.
The authors of \cite{busari18} assess the network performances in terms of the cell capacity, which can be expressed as
\begin{align}
    C=B\mbox{log}_2( & 1+P_{TX}+G_{TX}+G_{RX}-PL-SF-\sum_{n=1}^{N_{int}}I_n\nonumber\\
    & -N_0-10\mbox{log}_{10} B-NF),
\end{align}
where $B$ is the effective bandwidth, $P_{TX}$ is the transmit power, $G_{TX}$ is the transmitter gain, $G_{RX}$ is the receiver gain, $PL$ is the path loss, $SF$ is the shadow fading, $N_{int}$ is the number of non-target links operating in the same frequency band, $I_n$ is the interference caused by the $n$th non-target link ($n=1,2,...,N_{int}$), $N_0$ is the noise power density, and $NF$ is the receiver noise figure.
It is shown in \cite{busari18} that the network performances do not scale proportionally with the increase in the effective bandwidth due to the larger noise induced by the larger bandwidth, and that the throughput of indoor users is significantly degraded compared with that of outdoor users because of the indoor and penetration losses \cite{yang15,mumtaz16}.
Furthermore, it is expected that the issue of coexistence of mmWave, microwave, and terahertz communications will rise, considering the growing interests toward terahertz bands \cite{akyildiz14,mumtaz17}.

In \cite{semiari19}, a heterogeneous network which integrates both mmWave and microwave tiers to enable joint enhanced mobile broadband (eMBB) \cite{carvalho17} and ultra-reliable low latency communications (URLLC) \cite{popovski14} is studied.
The authors of \cite{semiari19} propose three frameworks as a guideline.
The first one focuses on the integrated radio interface and frame structures, featuring the medium access control (MAC) layer integration for fast scheduling, the packet data convergence protocol (PDCP) layer integration for reliability maximization, and flexible frame structures for latency reduction \cite{semiari19}.
The second one focuses on multiple access and resource allocation, featuring microwave channel measurements and localization for overhead and delay minimization during beam training at mmWave radio access technologies (RATs), control-plane and user-plane separation, joint uplink-downlink traffic management and load balancing, and fast and reliable backhaul connectivity \cite{semiari19}.
The third one focuses on mobility management, featuring higher speeds of mmWave communications for mobile users, reduction of handover failures and traffic management for reliability enhancement, and data caching through mmWave RATs.

\subsection{Radio Architecture for Coexistence}
In order to allow the simultaneous operations of mmWave and microwave services, advanced radio architectures have been proposed to enable the coexistence of communications in both frequency bands.

Since optical fibers are proper media to offer a sufficient bandwidth \cite{green04}, the optical communications with radio over fiber (RoF) are considered as a solution to the coexistence of both mmWave and microwave services \cite{liu13,zhu13,dat14}.

In \cite{liu13b} and \cite{chang13}, the authors propose an access architecture adopting analog RoF, which enables the functional simplification of remote antenna units (RAUs) and the multi-service coexistence across different frequency bands, and wavelength division multiplexing (WDM) technologies, which provide flexibility to backhaul networks.

In \cite{dat16}, a mobile fronthaul architecture featuring simultaneous transmission of both mmWave and microwave signals is proposed.
In this architecture, the local oscillator signal generated from an optical mmWave signal generator is combined with the laser diode-modulated microwave signal using the optical modulation in \cite{kanno12}.
Then, the combined signal goes through the intermediate-frequency-over-fiber (IFoF) system in \cite{dat16b} and is separated by an optical coupler.
With digital signal processing techniques, the microwave signal and down-converted mmWave signal are obtained.
Finally, the mmWave signal is recovered after the up-conversion in either an electrical or an optical manner.

In addition to the above schemes, there is also a research direction focusing on expanding the legacy microwave radio architecture in order to allow the coexistence of both mmWave and microwave signals.

In \cite{huang17b}, the authors propose a radio transceiver architecture, which is expanded from the legacy microwave radio scheme, for the coexistence of both mmWave and microwave systems.
Within the architecture in \cite{huang17b}, the common parts shared by both mmWave and microwave systems include the in-phase and quadrature (IQ) modulator/demodulator (MODEM), filters and amplifiers for baseband analog signals, and the frequency synthesizer as a local oscillator.
For the mmWave radio, it works in the time-division-multiplexing (TDD) mode \cite{levanen14} with both uplink and downlink sharing the same frequencies, and adopts the frequency synthesizer and multiplier for the necessary frequency up-conversion or down-conversion.
For the microwave radio, it works in the frequency-division-multiplexing (FDD) mode, and its signals are processed by direct-conversion.


\subsection{New Components for Coexistence}
In order to enable the coexistence of both mmWave and microwave technologies in the design of integrated circuits, new components are proposed as one of the solutions.

To better facilitate the transmission and reception of wireless communication systems across both mmWave and microwave bands, novel antenna design is required.

In \cite{ren18}, the authors propose a coexistent antenna structure, where two linear antenna arrays are adopted to compensate the path loss over the mmWave band \cite{zhao18}, while a $8\times 8$ MIMO system is constructed to achieve the sufficient data rate over the microwave band \cite{li16b,ban16}.

In \cite{zheng18}, the author proposes a dual-band antenna structure, where a substrate integrated waveguide (SIW) cross-slot antenna \cite{yang06} and an annular-ring antenna \cite{chew82} are used over the mmWave and microwave bands, respectively.
In addition, \cite{zheng18} adopts another smaller annular-ring antenna to address the  misalignment between the main lobe and broadside direction.

Besides antennas, advanced design of other components has been proposed to further allow the simultaneous operations of both mmWave and microwave communications.

In \cite{tsai10}, a wide-IF-bandwidth Gilbert-cell mixer \cite{gilbert68}, whose load stage is realized by an inductor-capacitor (LC) resonant circuit with a switchable capacitor array, in a complementary metal-oxide-semiconductor (CMOS) process is proposed.
With the proposed mixer in \cite{tsai10}, the mmWave siganls are down-converted to the IF band and can be transmitted through an IFoF system.

In \cite{ye18}, the authors propose a dual-band quadrature coupler configuration, where SIWs and microstrips are adopted for mmWave and microwave bands, respectively.
Suppose SIWs work with the guided wavelength $\lambda_g$ at the operating frequency $f_0$, there are two points that need to be considered in the design of \cite{ye18}.
First, microstrips in a higher-order operating mode affect SIWs.
To avoid this undesired effect, two slot apertures are deployed near the microstrips, since microstrips with slot apertures behave similar to a parallel resonant circuit with the band-stop property \cite{woo06}, and the length of both slot apertures should be
\begin{equation}
    L_{slot}=\frac{c_0}{2f_0}\sqrt{\frac{\epsilon+1}{2}},
\end{equation}
where $c_0$ is the light speed in free space and $\epsilon$ is the dielectric constant of substrates.
Second, the transmission between SIWs and microstrips requires a field matching \cite{garg13}, which can be obtained by setting the length of substrates as $\lambda_g$.
After addressing these two points, the design in \cite{ye18} can be utilized to enable the simultaneous operations over both mmWave and microwave bands.

\subsection{Resource Management for Coexistence}
With the rapid increase in new applications which are constrained by various quality-of-service (QoS) requirements, the resource management for proper user scheduling becomes an essential issue to be addressed in the coexistence of both mmWave and microwave communications \cite{wells10,park16}.

In \cite{semiari16} and \cite{semiari17}, the authors propose a context-aware dual-mode scheduling framework for joint mmWave and microwave resource allocation to user applications in user equipment.
With the goal of maximizing the number of satisfied user applications, the proposed framework in \cite{semiari16} and \cite{semiari17} uses a set of context information, including the channel state information (CSI) of user equipment, the delay tolerance and required load of user applications, and the uncertainty of mmWave channels, to perform user application selection and scheduling.
Within this framework, the mmWave transceivers adopt the antenna arrays to achieve a sufficient beamforming gain for LoS user equipment \cite{ghosh14}, while microwave transceivers are equipped with omni-directional antennas for low overhead and complexity \cite{ghosh09}.
For the joint context-aware user application selection and scheduling over the microwave band, it is formulated in \cite{semiari16} and \cite{semiari17} by the matching theory \cite{roth92,jorswieck11,semiari14,semiari15} as a matching game with externalities \cite{gu15} between user applications and microwave resource blocks.
With the proposed algorithm in \cite{semiari16} and \cite{semiari17}, it is guaranteed that this game will end up with a two-side stable matching between user applications and microwave resource blocks.
To enable the joint context-aware user application selection and scheduling over the mmWave band, the base stations need to have the information of each user application's LoS probability, which is obtained in \cite{semiari17} with a Q-learning (QL) method \cite{watkins92,simsek12,sutton18} where the user equipment monitors the successful LoS transmissions over time and sends the results back to the base stations.
Subsequently, the joint selection and scheduling is formulated in \cite{semiari17} as a 0-1 stochastic Knapsack optimization problem \cite{dean05}, which is solved in \cite{semiari17} by an iterative sorting algorithm.

In \cite{chergui19}, the authors propose a semi-blind classification algorithm for joint mmWave and microwave resource allocation in both uplink and downlink to a specific access point.
Initially, the QuaDRiGa simulator \cite{jaeckel14} is adopted to generate $M$ synthesized access points, with each point connected by $L$ links.
Then, the Rician $K$-factor \cite{atzeni18} and downlink reference signal receive power (RSRP) \cite{sesia11} of the $i$th access point ($i=1,2,...,M$), expressed as
\begin{equation}
    K_i=\frac{|h_{0,i}|^2}{\sum_{l=1}^{L-1}|h_{l,i}|^2},
\end{equation}
where $h_{0,i}$ and $h_{l,i},l=1,2,...,L-1$ are the only LoS path gain and the $l$th NLoS path gain of the $i$th access point, and $P_i$, respectively, are collected and subsequently projected into a 3D space with principal component analysis (PCA) \cite{tipping99}.
Finally, the projected Rician $K$-factor and downlink RSRP are used as the training data of a support vector machine (SVM) with nonlinear radial basis function (RBF) kernels \cite{cortes95}, and the trained SVM can be used for the prediction of joint mmWave and microwave resource management for a future access point.

\subsection{Coexistence in wireless local area networks (WLANs)}
\begin{figure*}[ht]
\centering
\includegraphics[width=16cm]{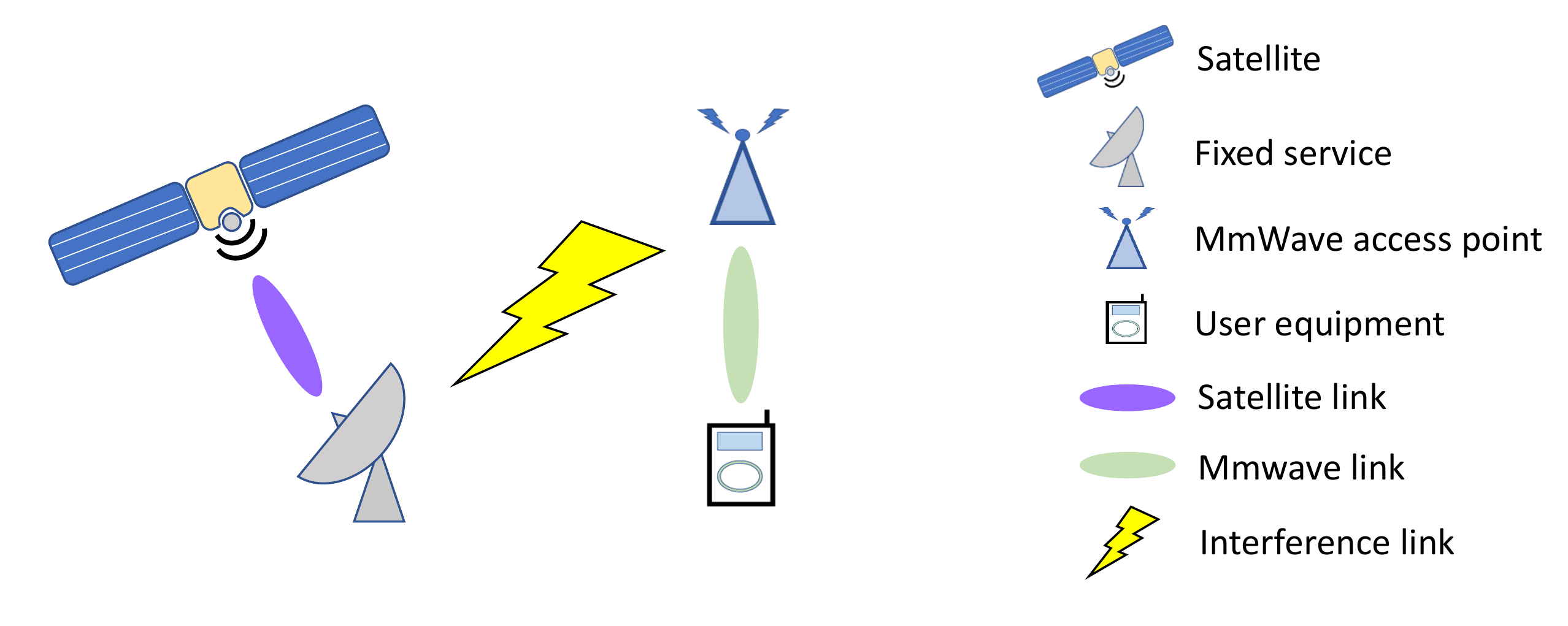}
\caption{Illustration of coexistence of mmWave communications and fixed services}
\label{fig:cat2}
\end{figure*}
Over the last few years, there is a growing demand of bandwidth-hungry applications, such as web conferencing and video streaming, which engender an explosion of traffic to WLANs.
In order to alleviate the heavy traffic in WLANs, the mmWave band plays a crucial role of sharing the workload of the microwave band, and therefore the coexistence of both mmWave and microwave bands becomes an essential issue to WLANs.

In \cite{park08}, it is envisioned that the integration of both mmWave and microwave bands will take place in either the link layer or the MAC layer of WLANs.
For the link layer, the handover process can be integrated over both mmWave and microwave bands with hooks defined to enable a fast context transfer between two radios.
Although the handover integration in the link layer will not cause service interruptions during the simultaneous operations of both bands, the latency may rise as a problem since the handover process takes a relatively long time.
For the MAC layer, there exist two types of integration, upper MAC integration and full MAC integration, depending on both lower MAC functions and upper MAC functions.
Lower MAC functions are based on hardware, requiring precise timing and rapid responses, while upper MAC functions are based on software, allowing rather loose timing requirements.
Accordingly, the integration is called upper MAC integration if only upper MAC functions are involved, and is called full MAC integration if both upper and lower MAC functions are involved.
Both types of MAC layer integration take a relatively short time, which will not cause a significant latency, but they might not allow the simultaneous operations of both mmWave and microwave services.

In \cite{semiari17b}, a MAC protocol where multi-band stations can adopt fast session transfers (FSTs) to offload traffic that causes excessive delays from the congested microwave band to the mmWave band in WLANs is proposed.
The scenario considered in \cite{semiari17b} begins with a station $S_1$ transmitting a packet to another station $S_2$ over the microwave band.
If the number of transmission failures of the packet reaches $m$, $S_1$ initiates an FST with $S_2$ with probability $\beta$.
Suppose the FST happens.
Initially, the MAC layer of $S_1$ sends an FST setup request to $S_2$, and then $S_2$ handshakes with $S_1$ for confirmation.
Subsequently, the mmWave MAC layer management entity (MLME) of $S_1$ initiates an FST acknowledgement request to $S_2$, and then $S_2$ sends an FST acknowledgement response to $S_1$.
At this moment, the FST is done and the packet transmission is ready to be reallocated to the mmWave band.
Finally, $S_1$ requests a contention-free time slot and transmit the packet during the time slot over the mmWave band.

In \cite{nguyen17}, the authors study the adoption of multipath transmission control protocol (MPTCP) \cite{ford12}, due to its switchover capability \cite{paasch12,chen13,lim14,nguyen14}, for the coexistence of both mmWave and microwave bands in WLANs.
Particularly, \cite{nguyen17} experiments with MPTCP in two modes: fullmesh and backup, and demonstrates that the backup mode of MPTCP can provide a sufficient throughput in the coexistence of both mmWave and microwave bands.

\section{Coexistence with Fixed Services}
\label{sec:FS}
In this section, we study the issues regarding the
coexistence of cellular networks and fixed services over the mmWave band.
An illustration of coexistence of mmWave communications and fixed services is shown in Fig. \ref{fig:cat2}, where additional interference links affect the operation of both cellular networks and fixed services.


\subsection{Coexistence with Fixed Satellite Services}
Over the past years, various fixed satellite services work steadily over the mmWave band \cite{apostolidis13,liolis13,ippolito17}.
However, the incoming mmWave communications operated by next-generation cellular networks and these fixed satellite services will interfere with each other.
As a result, the level of protection from the interference generated by the interferer for the victim needs to be determined in order to allow their coexistence.

In \cite{guidolin15,guidolin15b,guidolin15c}, the authors investigate the coexistence of both cellular networks and fixed satellite services over the mmWave band.
The log-scale interference generated by a base station to a fixed satellite service can be expressed as \cite{guidolin15,guidolin15b}
{\small
\begin{equation}
    I=P_{BS}+G_{omni}+10\mbox{log}(|\vv^T \vh_{FSS}|^2)+G_{FSS}(\phi)-L(d),
\end{equation}
}where $P_{BS}$ is the base station transmit power, $G_{omni}$ is the antenna gain without beamforming, $\vv$ is the beamforming precoding vector selected by the base station, $\vh_{FSS}$ is the channel matrix between the base station and the fixed satellite service, $d$ is the distance between the base station and fixed satellite service, $L(d)$ is the path loss at the distance $d$, and $G_{FSS}(\phi)$ is the fixed satellite service antenna gain in the direction $\phi$ and can be expressed as
\begin{equation}
G_{FSS}(\phi) = \left\{\begin{array}{ll}
G_{max}, & 0^{\circ}<\phi<1^{\circ}\\
32-25\mbox{log}\phi, & 1^{\circ}\leq\phi<48^{\circ}\\
-10, & 48^{\circ}\leq\phi<180^{\circ}
\end{array}\right.,
\label{eq:gmn}
\end{equation}
where $G_{max}$ is the fixed satellite service gain in the main beam axis, $\phi=\mbox{cos}^{-1}(\mbox{cos}(\alpha)\mbox{cos}(\theta)\mbox{cos}(\epsilon)+\mbox{sin}(\alpha)\mbox{sin}(\epsilon))$, where $\alpha$ is the fixed satellite service elevation angle, $\theta$ is the base station azimuth with regard to the main lobe of the fixed satellite service receiver, and $\epsilon=\frac{h_t-h_s}{d}-\frac{d}{2r}$, where $h_t$ and $h_s$ are the height of the base station and fixed satellite service, respectively, and $r$ is the effective earth radius.

Specifically, \cite{guidolin15} proposes a cooperative scheduling algorithm for base stations to control the level of interference received by fixed satellite services in downlink \cite{bai14,hur15} by modeling the coexistence as a potential game \cite{monderer96}, which can be used to tackle different tasks in wireless networks \cite{nie06,buzzi12,chen13b}.
There are three versions of the proposed algorithm in \cite{guidolin15}.
The first one is aimed at maximizing the mean spectral efficiency of users within the coexistence, the second one is aimed at minimizing the interference received by fixed satellite receivers, and the third one is aimed at balancing between the mean spectral efficiency of users and the interference received by fixed satellite receivers.
With the proposed algorithm in \cite{guidolin15}, it can be shown that the considered potential game becomes a specific one with a Nash equilibrium solution \cite{monderer96,nie06}.

Moreover, \cite{guidolin15b} considers three scenarios in the coexistence of both cellular networks and fixed satellite services.
In the first scenario, where a single omni-directional base station is evaluated, it is shown that a larger fixed satellite service elevation angle $\alpha$ leads to less interference received by fixed satellite services.
In the second scenario, where multiple omni-directional base stations whose parameter setting follows \cite{rangan14} are evaluated, it is shown that a larger distance $d$ between base stations and fixed satellite services, and a larger fixed satellite service elevation angle $\alpha$ lead to less interference received by fixed satellite services.
In the third scenario, where multiple directional base stations equipped with multiple antennas and RF beamforming are evaluated, it is shown that RF beamforming and a larger antenna array in base stations, and a larger distance between base stations lead to less interference received by fixed satellite services.

Furthermore, \cite{guidolin15c} studies the effects of the coexistence of base stations and a specific fixed satellite service on the performances of user equipment.
In the scenario considered in \cite{guidolin15c}, there is no cooperation between base stations, and the path loss model in \cite{hur15} and the single-input-single-output (SISO) long-term evolution (LTE) in \cite{mogensen07} are adopted.
Suppose the user equipment is served by the base station indexed as $j$.
Then, the performances of user equipment are evaluated in terms of the capacity, which is expressed as
\begin{equation}
    C=B\mbox{log}(1+\frac{P_{BS}+G_{BS}-L(d_j)}{N+I_j}),
\end{equation}
where $B$ is the effective bandwidth, $N$ is the noise power density, $d_j$ is the distance between the user equipment and base station indexed as $j$, and $I_j$ is the interference, which can be expressed as
{\small
\begin{equation}
    I_j=P_{FSS}+G_{FSS}(\phi)-L(d_{FSS})+\sum_{n\neq j} P_{BS}+G_{BS}-L(d_n),
\end{equation}
}where $P_{FSS}$ is the fixed satellite service transmit power, and $d_{FSS}$ and $d_n$ are the distances between the user equipment and the fixed satellite service and between the user equipment and the base station indexed as $n$, respectively.
Note that the other notations follow \cite{guidolin15} and \cite{guidolin15b}.
It is found in \cite{guidolin15c} that the performances of user equipment can be enhanced with beamforming and an increase in the antenna number, the base station density, the fixed satellite service elevation angle, and the distance between the fixed satellite service and its nearest neighboring base station.

In \cite{kim17}, the authors investigate the interference from access points to space stations and from earth stations to access points within the coexistence of cellular networks and fixed satellite services over the mmWave band.
Suppose the distribution of user equipment follows a Poisson point process in a sector region \cite{chiu13}.
Note that the NLoS channel conditions for the interference from access points to space stations include a clutter loss, which is one of the sources of the diffraction loss \cite{anderson03}.
Denote the set of 5G sectors as $\cS_{5s}$, which contains $\bN[\cS_{5s}]$ sectors.
Then, the interference from access points to space stations and from earth stations to access points can be expressed as \cite{kim17}
\begin{equation}
    I_{aggr}(\bN[\cS_{5s}])=I_{5g}\times\bN[\cS_{5s}]
\end{equation}
and
\begin{equation}
    I_{es}=\frac{P_{T,es}G_{es,a}G_{ap,a}(x_{ue})G_{ap,e}(x_{ue})}{\mbox{PL}_{es\rightarrow ap}},
\end{equation}
respectively, where
{\scriptsize
\begin{equation}
    I_{5g}=\frac{1}{|{\cR_k}^2|}\int_{x_{ue}\in{\cR_k}^2}\frac{P_{T,ap}G_{ap,a}(x_{ue})G_{ap,e}(x_{ue})G_{ss,3db}}{\mbox{PL}_{ap\rightarrow ss}}dx_{ue},
\end{equation}
}where ${\cR_k}^2$ and $|{\cR_k}^2|$ are the region and area of a sector, respectively, $P_{T,ap}$ and $P_{T,es}$ are the transmit power of the access point and the earth station, respectively, $\mbox{PL}_{ap\rightarrow ss}$ and $\mbox{PL}_{es\rightarrow ap}$ are the path loss between the access point and the space station and between the earth station and the access point, respectively, $G_{ap,a}(x_{ue})$ and $G_{ap,e}(x_{ue})$ are the azimuth and elevation beamforming gains of a downlink transmission to user equipment with position $x_{ue}$ in the direction toward the space station, respectively, $G_{ss,3db}$ is the beamforming gain of the space station receiver antenna within its 3dB-contour, and $G_{es,a}$ is the azimuth of the earth station.

In \cite{lin18}, the authors propose a physical layer security framework for the coexistence of cellular networks and fixed satellite services over the mmWave band, extending from \cite{sharma13,vassaki13,an16,an16b,an16c,an17}.
Within the framework, the LoS path loss model in \cite{rappaport96} and the passive eavesdroppers characterized in \cite{zou14} are adopted.
Accordingly, the scenario becomes a constrained optimization problem to maximize the worst-case achievable secrecy rate of secondary users while satisfying the constraints of a base station transmit power budget and an allowable interference level to primary users, given that either coordinated or uncoordinated eavesdroppers exist.
For the presence of coordinated eavesdroppers, \cite{lin18} transforms the original problem into a min-max one and proposes a heuristic beamforming scheme to solve it.
For the presence of uncoordinated eavesdroppers, \cite{lin18} introduces an auxiliary variable to convert the original problem into a semi-definite programming one with rank-one constraints and proposes an iterative penalty function based algorithm to solve it.

In \cite{meng19}, the authors study the coexistence of cellular networks and fixed satellite services at 40 GHz.
In the scenario considered in \cite{meng19}, the number of base stations, which adopt massive MIMO systems \cite{hong14} with the channel models in \cite{han19}, is
\begin{equation}
    N=S_{urb}R_{a_{urb}}D_{s_{urb}}+S_{sub}R_{a_{sub}}D_{s_{sub}},
\end{equation}
where $S_{urb}$ and $S_{sub}$ are the sizes of urban and suburban areas, respectively, $R_{a_{urb}}$ and $R_{a_{sub}}$ are the ratios of hotspot areas to urban and suburban areas, respectively, and $D_{s_{urb}}$ and $D_{s_{sub}}$ are the hotspot densities of urban and suburban hotspot areas, respectively.
It is found in \cite{meng19} that the interference from base stations to fixed satellite services can be mitigated with massive MIMO due to the resulting narrower beams and more space diversity.

    
\subsection{Coexistence with Other Incumbents}
Besides fixed satellite services, there are other incumbents working over the mmWave band.
Therefore, the coexistence of next-generation cellular networks and these incumbents in mmWave frequencies needs to be considered as well.

In particular, there is a growing interest in the coexistence of cellular networks and existing incumbents over the band beyond 70 GHz \cite{federal16}, since this band is licensed worldwide and features a large available spectrum \cite{hattab18}.
Over this band, the interference from the interferer to the victim depend on three factors: the path loss, the equivalent isotropic radiated power (EIRP) from the interferer, and the effective antenna gain at the victim \cite{kim17,hattab17,hattab19}.

Generally, the existing interference mitigation techniques can be classified as two categories, including active mitigation and passive mitigation.
For the active mitigation, probes installed near the victim monitor the level of interference caused by each of the cells and require the cells generating an excess level of interference offload part of work to other cells \cite{hattab18}.
For the passive mitigation, the common approaches adopted include angular exclusion zones \cite{kim17,hattab18b,hattab19}, which limit the use of beams within some sectors in the angular domain, and power control \cite{hattab19}, which adjust the interferer transmit power.

In addition, there is other research focusing on the coexistence at different frequencies.

In \cite{kim15}, the authors investigate the downlink and uplink interference between base stations and mobile stations to fixed services at 39 GHz.
According to \cite{kim15}, the downlink and uplink interference between a base station $i$ and its associated mobile station $j$ to a fixed service $k$ can be expressed as
\begin{equation}
    I_{d,ijk}=P_i+G(\phi^+,\theta^+)+G(\phi^*,\theta^*)-L(d_{ik})
\end{equation}
and
\begin{equation}
    I_{u,ijk}=P_j+G(0,0)+G(\phi^*,\theta^*)-L(d_{jk}),
\end{equation}
respectively, where $P_i$ and $P_j$ are the transmit power of the base station $i$ and the mobile station $j$, respectively, 
$\phi^+$ and $\theta^+$ are the azimuth and the elevation angle between the transmit beam of the base station $i$ and the fixed service $k$, respectively, $\phi^*$ and $\theta^*$ are the azimuth and the elevation angle between the receive beam of the fixed service $k$ and the base station $i$, respectively, $L(d_{ik})$ and $L(d_{jk})$ are the attenuation regarding the distance between the base station $i$ and fixed service $k$ and the distance between the mobile station $j$ and fixed service $k$, respectively,
and $G(\phi,\theta)$ is the receiver antenna gain of the fixed service $k$ at the azimuth $\phi$ and elevation angle $\theta$.

In \cite{choi18}, the coexistence of both ITS and other wireless services over the 60 GHz band is studied.
Specifically, \cite{choi18} evaluates the required minimum coupling loss (MCL) \cite{administrations99}, which is the required minimum path loss to suppress the interference from the interferer to the victim to an acceptable level of the interference-to-noise ratio ($\frac{I}{N}$) and carrier-to-interference ratio ($\frac{C}{I}$), expressed as
\begin{equation}
MCL(\frac{I}{N})=P_{int}-L_{vic}-\frac{I}{N}-N
\label{eq:MCL_IN}
\end{equation}
and
\begin{equation}
MCL(\frac{C}{I})=P_{int}+L_{vic}+\frac{C}{I}-ST_{vic},
\label{eq:MCL_CI}
\end{equation}
respectively, where $P_{int}$ is the EIRP of the interferer, $L_{vic}$ and $ST_{vic}$ are the sidelobe attenuation and antenna sensitivity of the victim, respectively.

\section{Coexistence with NOMA}
\label{sec:NOMA}
\begin{figure}[ht]
\centering
\includegraphics[width=8.5cm]{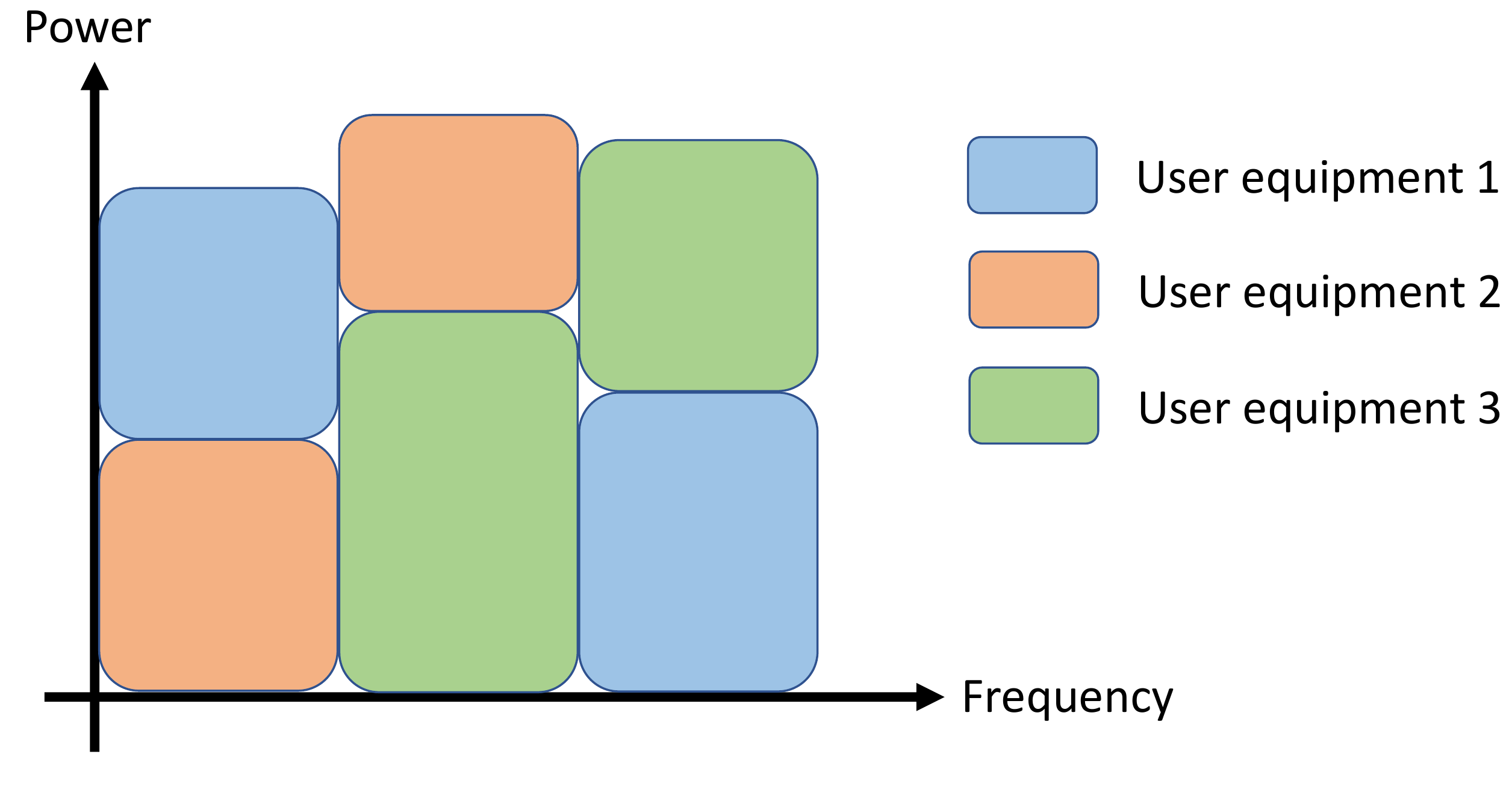}
\caption{Illustration of principles of NOMA}
\label{fig:cat3}
\end{figure}
In this section, we focus on the issues regarding the coexistence of mmWave communications and NOMA in next-generation cellular networks.
An illustration of the principles of NOMA is shown in Fig. \ref{fig:cat3}, where different user equipment can occupy the same frequency bin if allocated different power.


In \cite{ding17} and \cite{ding17b}, the authors adopt the random beamforming in the downlink transmission with the coexistence of mmWave communications and NOMA.
For the system model, \cite{ding17} and \cite{ding17b} assume that the user distribution in a disc area follows a homogeneous Poisson point process \cite{haenggi12}, and use a LoS-dominant channel model \cite{rappaport12,lee16} with the LoS probability derived in \cite{andrews10} and \cite{thornburg16}.
In addition, \cite{ding17} and \cite{ding17b} utilize the analog precoding for mmWave communications, the two-user case for NOMA \cite{sun15,ding16}, and the beamformer in \cite{lee16} and \cite{lee16b}, which is a special case of \cite{gao16}.
Accordingly, \cite{ding17} and \cite{ding17b} analyze the single-beam and multiple-beam cases in terms of the sum rate and outage probability, and propose two random beamforming schemes with limited feedback in order to reduce the system overhead.
The first scheme considers only the distances between base stations and users as the available information, while the second scheme further restricts the available information to one bit representing the channel quality, according to a predetermined threshold, sent from users to base stations.

In \cite{zhou18} and \cite{kusaladharma18}, the uplink and downlink transmissions of joint mmWave and NOMA systems under more realistic conditions are studied.
It is shown in \cite{zhou18} and \cite{kusaladharma18} that the performances of these systems can be enhanced with a denser deployment of base stations. 

In \cite{zhang17b}, the multicasting, which is widely adopted in wireless networks \cite{lin14b,kim16,fuente16,condoluci16,araniti17}, within the coexistence of mmWave communications and NOMA is studied.
Specifically, \cite{zhang17b} demonstrates that the multicasting spectrum efficiency in mmWave communications can be enhanced with NOMA, provides an analysis of the corresponding coverage probability, sum multicast rate, and average number of served users, and proposes cooperative algorithms to improve the system performances.

In \cite{morgado17}, the authors investigate the effect of joint NOMA and orthogonal frequency division multiple access (OFDMA) resource allocation to user equipment with mmWave communications on the cell capacity.
Note that in a two-user case (where two users are in the same cell), with $UE_1$ and $UE_2$ representing the user closer to the base station and cell edge, respectively, the power allocated to $UE_1$ and $UE_2$ given a total power budget $P_{total}$ with NOMA can be expressed as \cite{zhu15}
\begin{equation}
    P_{UE_1}=\frac{\sqrt{1+\mbox{SNR}_{UE_2}}-1}{\mbox{SNR}_{UE_2}}\times P_{total}
    \label{eq:UE1}
\end{equation}
and
\begin{equation}
     P_{UE_2}=P_{total}-P_{UE_1},
     \label{eq:UE2}
\end{equation}
respectively, where $\mbox{SNR}_{UE_1}$ and $\mbox{SNR}_{UE_2}$ are the signal-to-noise ratio (SNR) of users $UE_1$ and $UE_2$, respectively.
It is found in \cite{morgado17} that the cell capacity is maximized when NOMA is used for all user equipment.

\section{Other Coexistence Issues}
\label{sec:other}
In this section, we introduce the other coexistence issues of mmWave communications.

With recent advancements in mmWave communications, it is expected that various services will be operated over the mmWave band in the near future.
Therefore, the coexistence of different services at mmWave frequencies becomes an essential task that needs to be addressed in the design of next-generation communications.

In \cite{nekovee16} and \cite{nekovee17}, the authors investigate the coexistence of multiple RATs, which transmit signals via beamforming, within mmWave networks.
During the scheduling for beamforming in \cite{nekovee16} and \cite{nekovee17}, a beam sequence with indices of beams employed at different time slots is formed.
Accordingly, the scheduling for an optimal beam sequence to maximize the spectral efficiency of the whole network is a combinatorial optimization problem, which is NP-hard and computationally intractable \cite{trevisan11}.
In order to deal with this problem, \cite{nekovee16} and \cite{nekovee17} propose two algorithms.
The first one is called distributed greedy scheduling, which maximizes the individual utility function of each RAT with a block-coordinated optimization algorithm \cite{bazaraa13}.
The second one is called distributed learning scheduling, which allocates an initial probability to each possible sequence, and update the probability and utility function of each sequence iteratively until the number of iterations hits a predetermined threshold.

In \cite{anjum17}, the coexistence of multiple wireless body area networks (WBANs) at mmWave frequencies is studied.
Inspired by the therapeutic applications of mmWave \cite{chahat13,sarimin14}, the mmWave WBANs are proposed as a solution to the industry of telemedicine and e-health; however, these networks suffer from a severe interference in densely populated areas \cite{fang16}.
In order to combat this drawback, \cite{anjum17} formulates the coexistence issue as a non-cooperative and distributed game, where each network selects an optimal transmission power to maximize its signal-to-interference-plus-noise ratio (SINR), and derives the Nash equilibrium solution \cite{mackenzie06} to this game.
Moreover, \cite{anjum17} proposes a pricing policy to further improve the Nash equilibrium solution's efficiency in terms of the Pareto optimality and social optimality \cite{bacci15}.

In \cite{gupta16}, the coexistence of multiple operators, with each operator having its own spectrum license with a fixed bandwidth, over the mmWave band is investigated.
Within the scenario considered in \cite{gupta16}, the distribution of the base stations, which are equipped with steerable antennas whose radiation pattern follows \cite{bai15}, of each operator follows a Poisson point process, which can be divided into two Poisson point processes containing the base stations with LoS links and with NLoS links, respectively, to a specific user \cite{chiu13}.
Subsequently, the SINR and rate coverage probability of each operator for the specific user are explicitly derived in \cite{gupta16}.
In addition, it is shown in \cite{gupta16} that the optimal level of spectrum sharing depends on the target system rate.

In \cite{deng18}, the authors study the effect of heterogeneous antenna arrays on the coexistence of multiple devices, which enable D2D communications, over the mmWave band.
For the system model, \cite{deng18} assumes that the D2D users form a $K$-tier Poisson bipolar network \cite{haenggi12,haenggi14}, where the gain function of antenna arrays with a specific pattern in any tier follows \cite{balanis16}, and adopts the LoS model in \cite{singh15} and the cosine antenna pattern in \cite{yu17} to reflect the blockage effect.
Accordingly, \cite{deng18} proposes a mathematical framework, where the statistics of the interference caused by heterogeneous antenna arrays with different beams can be derived as closed-form expressions and further approximated by a mixture of the inverse gamma and log-normal distributions, and provides analytical expressions of both SINR and rate, which can be bounded or approximated for a tractable computation, of D2D networks with mmWave communications.


In \cite{mishra19}, the authors review the spectrum sharing among multiple radar systems over the mmWave band for mmWave joint radar communications, whose feasibility has been demonstrated in \cite{cohen18}, from a signal processing perspective.
Generally, the effective spectral efficiency of mmWave joint radar communications depends on the receiver implementation \cite{shi04,takizawa12,liu13c}.
In order to achieve a better interference management for mmWave joint radar communications, several schemes have been proposed focusing on either transmitters (e.g., \cite{mahal17,cui18}) or receivers (e.g., \cite{geng18,ayyar19}).

Besides the above issues, several other issues regarding mmWave communications have been studied in the literature as well.

In \cite{olmos08} and \cite{won10}, two architectures that enable the coexistence of wireless mmWave WDM-RoF and wired baseband signals are proposed.
Specifically, \cite{olmos08} and \cite{won10} adopt the optical-frequency interleaving in \cite{kuri07} and a reflective semiconductor optical amplifier, respectively, as the key technologies for the coexistence.

In \cite{yadav18}, the coexistence of mmWave communications and other technologies is studied.
Initially, \cite{yadav18} reviews several coexistence scenarios that involve mmWave communications and another single technology, e.g. the coexistence of mmWave communications and massive MIMO in \cite{bogale16}, and the coexistence of mmWave communications and ultra-dense networks in \cite{feng17}.
Then, \cite{yadav18} considers the coexistence of mmWave communications and other multiple technologies, including ultra-dense networks, optical wireless communications, massive MIMO, NOMA, and full duplex with an analysis, where mmWave communications generate a low inter-cell co-channel interference, ultra-dense networks reduce the access distance between users and base stations, optical wireless communications are suitable for indoor users, massive MIMO and NOMA can be combined for both multi-connectivity and improved power allocation, and full duplex enhances the spectral efficiency.


\section{Numerical evaluations}
\label{sec:simu}
In this section, we execute two evaluations, which are inspired by the literature, to simulate the coexistence issues of broadband mmWave communications and provide the numerical results for a further analysis.

The first evaluation studies the coexistence of ITS and other wireless services over the mmWave band, while the second evaluation studies the coexistence of mmWave communications and NOMA.


\subsection{Parameter Settings}
For the following two evaluations, the path loss (in dB) is expressed as
\begin{equation}
    PL=32.45+20\mbox{log}_{10}f+20\mbox{log}_{10}d,
    \label{eq:PL}
\end{equation}
where $f$ is the frequency (in MHz) and $d$ is the distance (in km).

In the first evaluation, we investigate the relation between i) the required protection distance $d$ between the interferer and victim and ii) the acceptable level of interference $\frac{I}{N}$ or $\frac{C}{I}$ for the victim with two cases within the coexistence of ITS and a wireless service.

We follow the parameter settings in \cite{administrations09} for this evaluation.
The frequency is set as $f=64$ GHz.
For the first case, we consider the scenario where ITS is the interferer and the wireless service is the victim, and the interference-to-noise ratio $\frac{I}{N}$ is adopted as the interference criterion for the victim.
The EIRP of the interferer is set as $P_{int}=-29$ dBm, and the sidelobe attenuation and noise power density of the victim are set as $L_{vic}=25$ dB and $N=-123$ dBm, respectively, in this case.
Besides, the interference-to-noise ratio $\frac{I}{N}$ is tested over $\{-40,-39,...,0\}$ dB in this case.
For the second case, we consider the scenario where the wireless service is the interferer and ITS is the victim, and the carrier-to-interference ratio $\frac{C}{I}$ is adopted as the interference criterion for the victim.
The EIRP of the interferer is set as $P_{int}=31$ dBm, and the sidelobe attenuation and antenna sensitivity of the victim are set as $L_{vic}=3$ dB and $ST=-86$ dBm, respectively, in this case.
Besides, the carrier-to-interference ratio $\frac{C}{I}$ is tested over $\{0,1,...,40\}$ dB in this case.
Note that the MCL in terms of the maximum acceptable $\frac{I}{N}$ and the minimum acceptable $\frac{C}{I}$, $MCL(\frac{I}{N})$ and $MCL(\frac{C}{I})$, can be computed with (\ref{eq:MCL_IN}) and (\ref{eq:MCL_CI}), respectively, and the protection distance $d$ can be obtained with (\ref{eq:PL}), where we substitute $MCL(\frac{I}{N})$ or $MCL(\frac{C}{I})$ for $PL$.

In the second evaluation, we consider a two-user case, with $UE_1$ and $UE_2$ representing the user (with a unity user equipment gain) closer to the base station and cell edge, respectively, within the coexistence of mmWave communications and NOMA, and investigate the relation between i) the capacity $C$ and ii) the distance to the base station $d$ of $UE_1$.




The carrier frequencies, effective bandwidth, noise power density, and base station transmit power are set as $f=\{30,70\}$ GHz, $B=5$ MHz, $N=-174$ dBm, and $P_{total}=40$ dBm.
Besides, the distance to base station $d$ is tested over $\{1,2,...,20\}$ km in this evaluation.
Note that the capacity (in bits/s) of $UE_1$, whose received interference is assumed to be perfectly cancelled, is expressed as
\begin{equation}
    C=B\mbox{log}_2(1+\mbox{SNR}_{UE_1})
\end{equation}
and the power allocation (in dBm) to $UE_1$ and $UE_2$, $P_{UE_1}$ and $P_{UE_2}$, can be computed with (\ref{eq:UE1}) and (\ref{eq:UE2}), respectively, iteratively until convergence, where $\mbox{SNR}_{UE_1}=P_{UE_1}-PL_{UE_1}-N$ and $\mbox{SNR}_{UE_2}=P_{UE_2}-PL_{UE_2}-N$, where $PL_{UE_1}$ and $PL_{UE_2}$ are the path loss of $UE_1$ and $UE_2$, respectively, obtained with (\ref{eq:PL}).
    


\subsection{Numerical Results}
\begin{figure}%
\centering
\subfigure[Case 1: the minimum protection distance $d$ with the corresponding maximum acceptable interference-to-noise ratio $I/N$]{%
\label{fig:Exp1_Case1}%
\includegraphics[width=8.5cm]{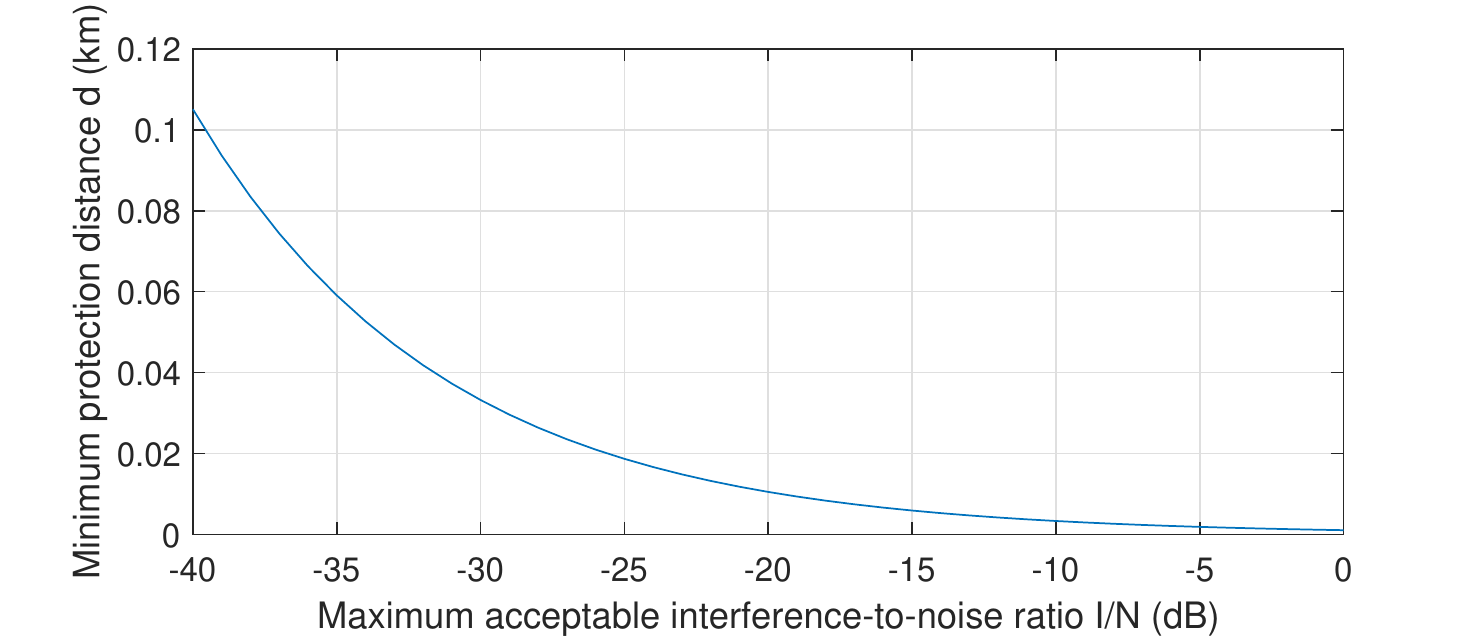}}%
\qquad
\subfigure[Case 2: the minimum protection distance $d$ with the corresponding minimum acceptable carrier-to-interference ratio $C/I$]{%
\label{fig:Exp1_Case2}%
\includegraphics[width=8.5cm]{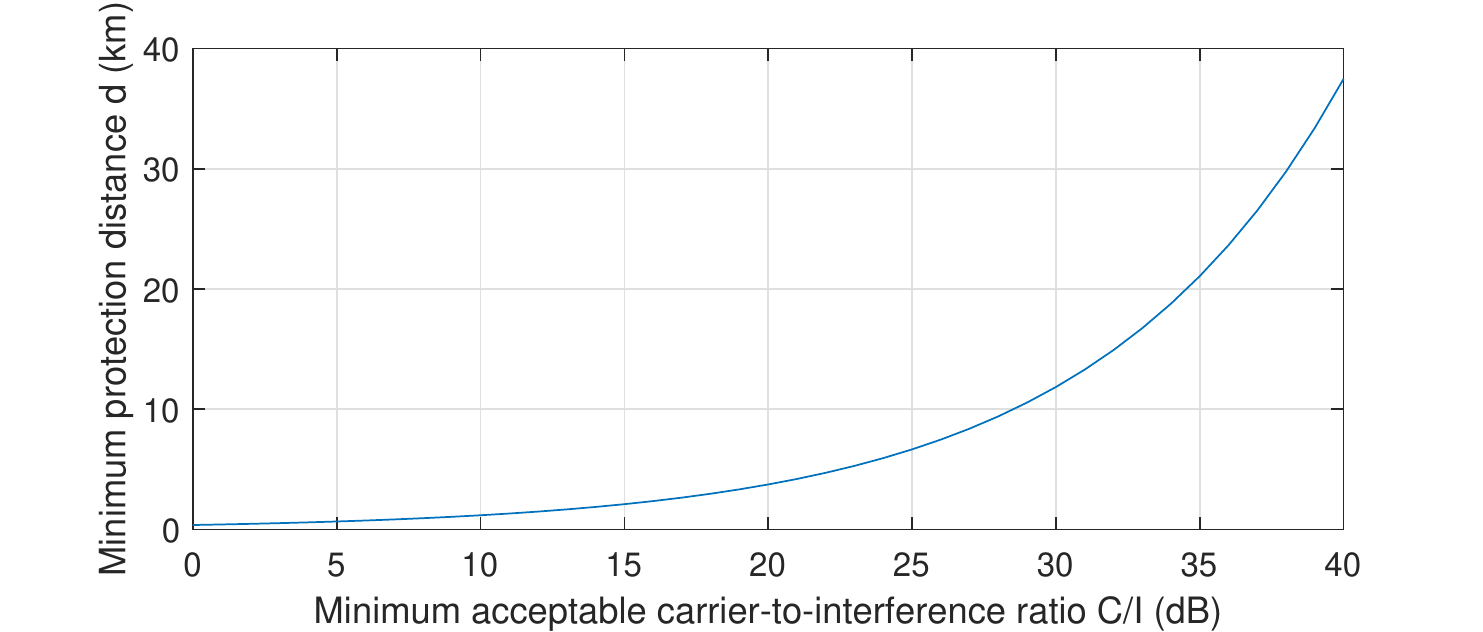}}%
\caption{Results of Evaluation 1}
\label{fig:Exp1}
\end{figure}

\begin{figure}[ht]
\centering
\includegraphics[width=8.5cm]{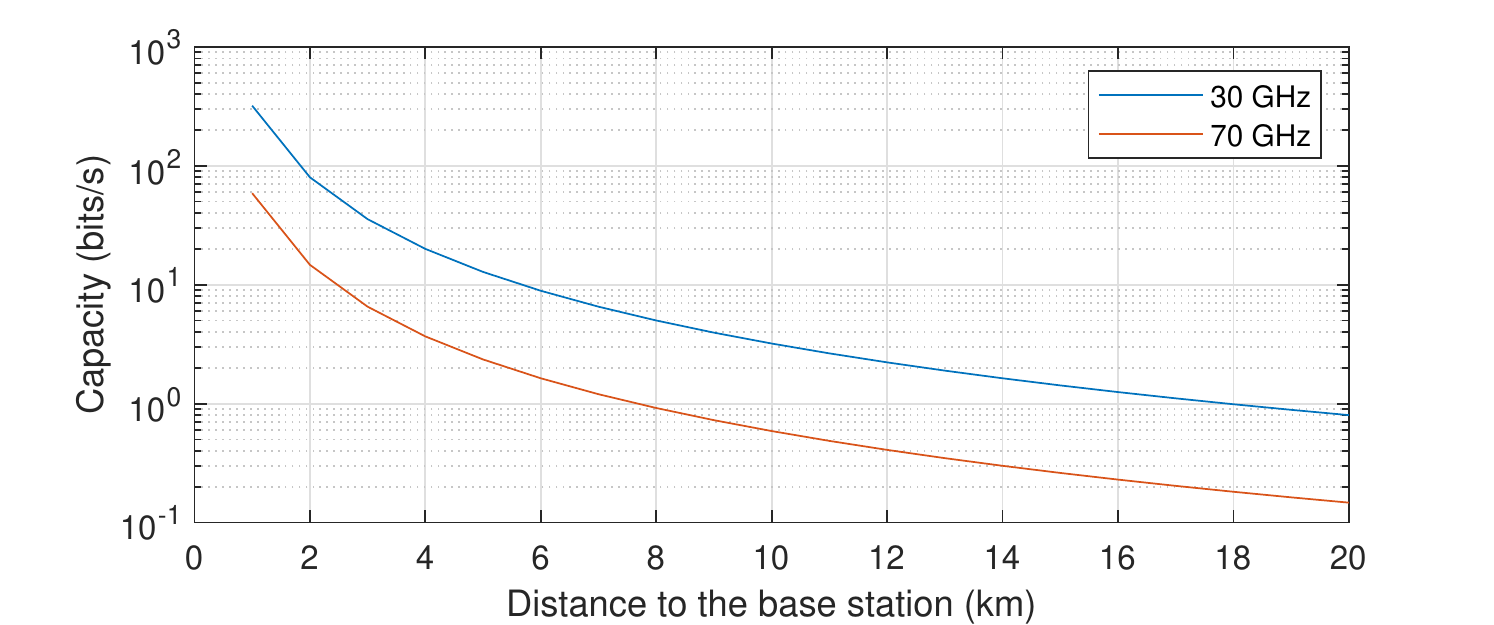}
\caption{Results of Evaluation 2}
\label{fig:Exp2}
\end{figure}
For the first evaluation, the results are shown in Fig. \ref{fig:Exp1}.
Specifically, the relations between the minimum protection distance $d$ and the maximum acceptable interference-to-noise ratio $\frac{I}{N}$ in the first case and between the minimum protection distance $d$ and the minimum acceptable carrier-to-interference ratio $\frac{C}{I}$ in the second case are shown in Figs. \ref{fig:Exp1_Case1} and \ref{fig:Exp1_Case2}, respectively.

According to the results in Fig. \ref{fig:Exp1}, it can be found that the minimum protection distance decreases with an increase in the maximum acceptable interference-to-noise ratio in a decreasing speed, while increases with an increase in the minimum acceptable carrier-to-interference ratio in an increasing speed.

For the second evaluation, the results are shown in Fig. \ref{fig:Exp2}, demonstrating the relation between the capacity and the distance to the base station of $UE_1$.

According to the results in Fig. \ref{fig:Exp2}, it can be observed that the capacity decreases with an increase in the distance to the base station in a decreasing speed.
In addition, it can be found that a lower frequency leads to an increased capacity.



\section{Challenges and Future Directions}
\label{sec:challenges}
The recent advancements in the coexistence issues of broadband mmWave communications, classified into four categories, have been systematically reviewed in Sec. \ref{sec:microwave}, \ref{sec:FS}, \ref{sec:NOMA}, and \ref{sec:other}, respectively.

In this section, we point out the challenges and future directions in each category.


\subsection{Coexistence with Microwave Communications}
In this category, novel structure design and cooperative schemes have been proposed for the coexistence of both mmWave and microwave communications.

However, there are still several challenges to be overcome in order to enhance their coexistence.
The high complexity for MAC protocols, mobility management, and beam tracking is a significant problem \cite{busari18,semiari19}.
Moreover, an efficient load balancing scheme is required in order to mitigate the high traffic in wireless networks \cite{semiari19}.
Furthermore, the cooperative caching and proper user association are needed for an efficient content caching \cite{zhu18}.

Therefore, the future directions include the design of low-complexity efficient control schemes and the effective strategies for content caching.



\subsection{Coexistence with Fixed Services}
In this category, the characterization and mitigation of interference within the coexistence of mmWave communications and fixed services have been discussed.

However, there are still some open challenges about their coexistence and the corresponding interference.
The spectrum availability within the coexistence has not been fully investigated.
Besides, existing interference mitigation techniques affect both uplink and downlink transmissions \cite{hattab18}.
Moreover, advanced antenna design and beamforming schemes are needed for better spectrum sharing.
In addition, the coexistence scenarios considered in the literature are based on rather unrealistic models \cite{guidolin15,guidolin15b,hattab18}.

Consequently, the future directions include the study of spectrum availability, novel interference mitigation techniques which strike a balance between the interference and transmission performances, advanced design of antennas and beamforming schemes, and the analysis based on more realistic models such as 3D stochastic processes.


\subsection{Coexistence with NOMA}
In this category, the combination of mmWave communications and NOMA for better spectrum sharing and power allocation within their coexistence has been investigated.

However, some challenges regarding their coexistence remain open and need to be addressed.
The applicability of NOMA to multi-cell mmWave networks where synchronization errors and inter-cell interference exist is still not clear \cite{morgado17}.
In addition, the coexistence analysis is mostly based on the perfect CSI, which is unrealistic.

As a result, the future directions include the feasibility of NOMA in multi-cell mmWave networks and the coexistence analysis based on the imperfect CSI.


\subsection{Other Coexistence Issues}
In this category, discrete topics from various perspectives on the coexistence issues of broadband mmWave communications have been studied.

However, there are still plenty of challenges to be overcome in these topics, and we mention some of them below.
The fairness among RATs within their coexistence over the mmWave band has not been fully considered in the literature \cite{nekovee16}.
Besides, novel multi-antenna techniques are required for an efficient license sharing \cite{gupta16}.

Therefore, the future directions include the fairness among RATs and new multi-antenna schemes over the mmWave band.



\section{Conclusions}
\label{sec:conclusion}
In this paper, we provide a systematic review of recent advancements in the coexistence issues of broadband mmWave communications, which are classified into four categories: coexistence with microwave communications, coexistence with fixed services, coexistence with NOMA, and other coexistence issues.
In addition, we present the results of numerical evaluations inspired by the literature for a further analysis and illustrate some challenges and future directions for each category of issues as a research roadmap.



%




\ifCLASSOPTIONcaptionsoff
  \newpage
\fi



%


\bibliographystyle{IEEEtran}
\bibliography{IEEEabrv,waveform}

\begin{thebibliography}{100}
\providecommand{\url}[1]{#1}
\csname url@samestyle\endcsname
\providecommand{\newblock}{\relax}
\providecommand{\bibinfo}[2]{#2}
\providecommand{\BIBentrySTDinterwordspacing}{\spaceskip=0pt\relax}
\providecommand{\BIBentryALTinterwordstretchfactor}{4}
\providecommand{\BIBentryALTinterwordspacing}{\spaceskip=\fontdimen2\font plus
\BIBentryALTinterwordstretchfactor\fontdimen3\font minus
  \fontdimen4\font\relax}
\providecommand{\BIBforeignlanguage}[2]{{%
\expandafter\ifx\csname l@#1\endcsname\relax
\typeout{** WARNING: IEEEtran.bst: No hyphenation pattern has been}%
\typeout{** loaded for the language `#1'. Using the pattern for}%
\typeout{** the default language instead.}%
\else
\language=\csname l@#1\endcsname
\fi
#2}}
\providecommand{\BIBdecl}{\relax}
\BIBdecl

\bibitem{xia12}
F.~Xia, L.~T. Yang, L.~Wang, and A.~Vinel, ``{Internet of things},''
  \emph{International Journal of Communication Systems}, vol.~25, no.~9, p.
  1101, 2012.

\bibitem{rangan14}
S.~{Rangan}, T.~S. {Rappaport}, and E.~{Erkip}, ``{Millimeter-Wave Cellular
  Wireless Networks: Potentials and Challenges},'' \emph{Proceedings of the
  IEEE}, vol. 102, no.~3, pp. 366--385, March 2014.

\bibitem{pi11}
Z.~{Pi} and F.~{Khan}, ``{An introduction to millimeter-wave mobile broadband
  systems},'' \emph{IEEE Communications Magazine}, vol.~49, no.~6, pp.
  101--107, June 2011.

\bibitem{administrations15}
WRC, ``{Studies on frequency-related matters for International Mobile
  Telecommunications identification including possible additional allocations
  to the mobile services on a primary basis in portion(s) of the frequency
  range between 24.25 and 86 GHz for the future development of International
  Mobile Telecommunications for 2020 and beyond},'' \emph{Resolution 238}, Nov
  2015.

\bibitem{li06}
Y.~Li and G.~Stuber, \emph{Orthogonal Frequency Division Multiplexing for
  Wireless Communications}.\hskip 1em plus 0.5em minus 0.4em\relax Springer,
  2006.

\bibitem{wells09}
J.~{Wells}, ``{Faster than fiber: The future of multi-G/s wireless},''
  \emph{IEEE Microwave Magazine}, vol.~10, no.~3, pp. 104--112, May 2009.

\bibitem{rappaport13}
T.~S. {Rappaport}, S.~{Sun}, R.~{Mayzus}, H.~{Zhao}, Y.~{Azar}, K.~{Wang},
  G.~N. {Wong}, J.~K. {Schulz}, M.~{Samimi}, and F.~{Gutierrez}, ``{Millimeter
  Wave Mobile Communications for 5G Cellular: It Will Work!}'' \emph{IEEE
  Access}, vol.~1, pp. 335--349, 2013.

\bibitem{kamel16}
M.~{Kamel}, W.~{Hamouda}, and A.~{Youssef}, ``{Ultra-Dense Networks: A
  Survey},'' \emph{IEEE Communications Surveys Tutorials}, vol.~18, no.~4, pp.
  2522--2545, Fourthquarter 2016.

\bibitem{saito13}
Y.~{Saito}, Y.~{Kishiyama}, A.~{Benjebbour}, T.~{Nakamura}, A.~{Li}, and
  K.~{Higuchi}, ``{Non-Orthogonal Multiple Access (NOMA) for Cellular Future
  Radio Access},'' in \emph{2013 IEEE 77th Vehicular Technology Conference (VTC
  Spring)}, June 2013, pp. 1--5.

\bibitem{dai15}
L.~{Dai}, B.~{Wang}, Y.~{Yuan}, S.~{Han}, C.~{I}, and Z.~{Wang},
  ``{Non-orthogonal multiple access for 5G: solutions, challenges,
  opportunities, and future research trends},'' \emph{IEEE Communications
  Magazine}, vol.~53, no.~9, pp. 74--81, Sep. 2015.

\bibitem{liu17}
Y.~{Liu}, Z.~{Qin}, M.~{Elkashlan}, Z.~{Ding}, A.~{Nallanathan}, and
  L.~{Hanzo}, ``{Nonorthogonal Multiple Access for 5G and Beyond},''
  \emph{Proceedings of the IEEE}, vol. 105, no.~12, pp. 2347--2381, Dec 2017.

\bibitem{ding17}
Z.~{Ding}, P.~{Fan}, and H.~V. {Poor}, ``{Random Beamforming in Millimeter-Wave
  NOMA Networks},'' \emph{IEEE Access}, vol.~5, pp. 7667--7681, 2017.

\bibitem{wang18}
L.~{Wang}, K.~{Wong}, S.~{Jin}, G.~{Zheng}, and R.~W. {Heath}, ``{A New Look at
  Physical Layer Security, Caching, and Wireless Energy Harvesting for
  Heterogeneous Ultra-Dense Networks},'' \emph{IEEE Communications Magazine},
  vol.~56, no.~6, pp. 49--55, June 2018.

\bibitem{zheng13}
G.~{Zheng}, I.~{Krikidis}, J.~{Li}, A.~P. {Petropulu}, and B.~{Ottersten},
  ``{Improving Physical Layer Secrecy Using Full-Duplex Jamming Receivers},''
  \emph{IEEE Transactions on Signal Processing}, vol.~61, no.~20, pp.
  4962--4974, Oct 2013.

\bibitem{mukherjee14}
A.~{Mukherjee}, S.~A.~A. {Fakoorian}, J.~{Huang}, and A.~L. {Swindlehurst},
  ``{Principles of Physical Layer Security in Multiuser Wireless Networks: A
  Survey},'' \emph{IEEE Communications Surveys Tutorials}, vol.~16, no.~3, pp.
  1550--1573, Third 2014.

\bibitem{yang15b}
N.~{Yang}, L.~{Wang}, G.~{Geraci}, M.~{Elkashlan}, J.~{Yuan}, and M.~D.
  {Renzo}, ``{Safeguarding 5G wireless communication networks using physical
  layer security},'' \emph{IEEE Communications Magazine}, vol.~53, no.~4, pp.
  20--27, April 2015.

\bibitem{zou15}
Y.~{Zou}, J.~{Zhu}, L.~{Yang}, Y.~{Liang}, and Y.~{Yao}, ``{Securing
  physical-layer communications for cognitive radio networks},'' \emph{IEEE
  Communications Magazine}, vol.~53, no.~9, pp. 48--54, Sep. 2015.

\bibitem{zhu17}
Y.~{Zhu}, L.~{Wang}, K.~{Wong}, and R.~W. {Heath}, ``{Secure Communications in
  Millimeter Wave Ad Hoc Networks},'' \emph{IEEE Transactions on Wireless
  Communications}, vol.~16, no.~5, pp. 3205--3217, May 2017.

\bibitem{andrews14}
J.~G. {Andrews}, S.~{Buzzi}, W.~{Choi}, S.~V. {Hanly}, A.~{Lozano}, A.~C.~K.
  {Soong}, and J.~C. {Zhang}, ``{What Will 5G Be?}'' \emph{IEEE Journal on
  Selected Areas in Communications}, vol.~32, no.~6, pp. 1065--1082, June 2014.

\bibitem{zhu18}
Y.~{Zhu}, G.~{Zheng}, L.~{Wang}, K.~{Wong}, and L.~{Zhao}, ``{Content Placement
  in Cache-Enabled Sub-6 GHz and Millimeter-Wave Multi-Antenna Dense Small Cell
  Networks},'' \emph{IEEE Transactions on Wireless Communications}, vol.~17,
  no.~5, pp. 2843--2856, May 2018.

\bibitem{blaszczyszyn15}
B.~{Blaszczyszyn} and A.~{Giovanidis}, ``{Optimal geographic caching in
  cellular networks},'' in \emph{2015 IEEE International Conference on
  Communications (ICC)}, June 2015, pp. 3358--3363.

\bibitem{bai15}
T.~{Bai} and R.~W. {Heath}, ``{Coverage and Rate Analysis for Millimeter-Wave
  Cellular Networks},'' \emph{IEEE Transactions on Wireless Communications},
  vol.~14, no.~2, pp. 1100--1114, Feb 2015.

\bibitem{park16}
J.~{Park}, S.~{Kim}, and J.~{Zander}, ``{Tractable Resource Management With
  Uplink Decoupled Millimeter-Wave Overlay in Ultra-Dense Cellular Networks},''
  \emph{IEEE Transactions on Wireless Communications}, vol.~15, no.~6, pp.
  4362--4379, June 2016.

\bibitem{wang17}
L.~{Wang} and K.~{Wong}, ``{Energy coverage in wireless powered sub-6 GHz and
  millimeter wave dense cellular networks},'' in \emph{2017 IEEE International
  Conference on Communications (ICC)}, May 2017, pp. 1--6.

\bibitem{An11}
S.~{An}, B.~{Lee}, and D.~{Shin}, ``{A Survey of Intelligent Transportation
  Systems},'' in \emph{2011 Third International Conference on Computational
  Intelligence, Communication Systems and Networks}, July 2011, pp. 332--337.

\bibitem{wu17}
C.~{Wu}, T.~{Yoshinaga}, and Y.~{Ji}, ``{Cooperative Content Delivery in
  Vehicular Networks with Integration of Sub-6 GHz and mmWave},'' in \emph{2017
  IEEE Globecom Workshops (GC Wkshps)}, Dec 2017, pp. 1--6.

\bibitem{zadeh88}
L.~A. {Zadeh}, ``{Fuzzy logic},'' \emph{Computer}, vol.~21, no.~4, pp. 83--93,
  April 1988.

\bibitem{anjinappa18}
C.~K. {Anjinappa} and I.~{Guvenc}, ``{Angular and Temporal Correlation of V2X
  Channels across Sub-6 GHz and mmWave Bands},'' in \emph{2018 IEEE
  International Conference on Communications Workshops (ICC Workshops)}, May
  2018, pp. 1--6.

\bibitem{rois16}
J.~G. {Rois}, B.~{Lorenzo}, F.~J. {González-Castaño}, and J.~C. {Burguillo},
  ``{Heterogeneous millimeter-wave/micro-wave architecture for 5G wireless
  access and backhauling},'' in \emph{2016 European Conference on Networks and
  Communications (EuCNC)}, June 2016, pp. 179--184.

\bibitem{kim14b}
Y.~{Kim}, H.~{Ji}, J.~{Lee}, Y.~{Nam}, B.~L. {Ng}, I.~{Tzanidis}, Y.~{Li}, and
  J.~{Zhang}, ``{Full dimension mimo (FD-MIMO): the next evolution of MIMO in
  LTE systems},'' \emph{IEEE Wireless Communications}, vol.~21, no.~2, pp.
  26--33, April 2014.

\bibitem{rois15}
J.~{García-Rois}, F.~{Gómez-Cuba}, M.~R. {Akdeniz}, F.~J.
  {González-Castaño}, J.~C. {Burguillo}, S.~{Rangan}, and B.~{Lorenzo}, ``{On
  the Analysis of Scheduling in Dynamic Duplex Multihop mmWave Cellular
  Systems},'' \emph{IEEE Transactions on Wireless Communications}, vol.~14,
  no.~11, pp. 6028--6042, Nov 2015.

\bibitem{pedersen11}
K.~I. {Pedersen}, F.~{Frederiksen}, C.~{Rosa}, H.~{Nguyen}, L.~G.~U. {Garcia},
  and Y.~{Wang}, ``{Carrier aggregation for LTE-advanced: functionality and
  performance aspects},'' \emph{IEEE Communications Magazine}, vol.~49, no.~6,
  pp. 89--95, June 2011.

\bibitem{deng17}
J.~{Deng}, O.~{Tirkkonen}, R.~{Freij-Hollanti}, T.~{Chen}, and N.~{Nikaein},
  ``{Resource Allocation and Interference Management for Opportunistic Relaying
  in Integrated mmWave/sub-6 GHz 5G Networks},'' \emph{IEEE Communications
  Magazine}, vol.~55, no.~6, pp. 94--101, June 2017.

\bibitem{ishii12}
H.~{Ishii}, Y.~{Kishiyama}, and H.~{Takahashi}, ``{A novel architecture for
  LTE-B :C-plane/U-plane split and Phantom Cell concept},'' in \emph{2012 IEEE
  Globecom Workshops}, Dec 2012, pp. 624--630.

\bibitem{osseiran14}
A.~{Osseiran}, F.~{Boccardi}, V.~{Braun}, K.~{Kusume}, P.~{Marsch},
  M.~{Maternia}, O.~{Queseth}, M.~{Schellmann}, H.~{Schotten}, H.~{Taoka},
  H.~{Tullberg}, M.~A. {Uusitalo}, B.~{Timus}, and M.~{Fallgren}, ``{Scenarios
  for 5G mobile and wireless communications: the vision of the METIS
  project},'' \emph{IEEE Communications Magazine}, vol.~52, no.~5, pp. 26--35,
  May 2014.

\bibitem{tiirola13}
E.~{Tiirola}, B.~{Raaf}, E.~{Lähetkangas}, I.~{Harjula}, and K.~{Pajukoski},
  ``{On the design of discovery patterns for half-duplex TDD nodes operating in
  frame-based systems},'' in \emph{2013 Future Network Mobile Summit}, July
  2013, pp. 1--9.

\bibitem{chaitin82}
G.~J. Chaitin, ``{Register allocation \& spilling via graph coloring},'' in
  \emph{ACM Sigplan Notices}, vol.~17, no.~6.\hskip 1em plus 0.5em minus
  0.4em\relax ACM, 1982, pp. 98--105.

\bibitem{busari18}
S.~A. {Busari}, S.~{Mumtaz}, S.~{Al-Rubaye}, and J.~{Rodriguez}, ``{5G
  Millimeter-Wave Mobile Broadband: Performance and Challenges},'' \emph{IEEE
  Communications Magazine}, vol.~56, no.~6, pp. 137--143, June 2018.

\bibitem{busari17}
S.~A. {Busari}, S.~{Mumtaz}, K.~M.~S. {Huq}, J.~{Rodriguez}, and H.~{Gacanin},
  ``{System-Level Performance Evaluation for 5G mmWave Cellular Network},'' in
  \emph{GLOBECOM 2017 - 2017 IEEE Global Communications Conference}, Dec 2017,
  pp. 1--7.

\bibitem{yang15}
G.~{Yang}, J.~{Du}, and M.~{Xiao}, ``{Maximum Throughput Path Selection With
  Random Blockage for Indoor 60 GHz Relay Networks},'' \emph{IEEE Transactions
  on Communications}, vol.~63, no.~10, pp. 3511--3524, Oct 2015.

\bibitem{mumtaz16}
S.~Mumtaz, J.~Rodriguez, and L.~Dai, \emph{{MmWave Massive MIMO: A Paradigm for
  5G}}.\hskip 1em plus 0.5em minus 0.4em\relax Academic Press, 2016.

\bibitem{akyildiz14}
I.~F. Akyildiz, J.~M. Jornet, and C.~Han, ``{Terahertz band: Next frontier for
  wireless communications},'' \emph{Physical Communication}, vol.~12, pp.
  16--32, 2014.

\bibitem{mumtaz17}
S.~Mumtaz, J.~M. Jornet, J.~Aulin, W.~H. Gerstacker, X.~Dong, and B.~Ai,
  ``{Terahertz communication for vehicular networks},'' \emph{IEEE Transactions
  on Vehicular Technology}, vol.~66, no.~7, pp. 5617--5625, 2017.

\bibitem{semiari19}
O.~{Semiari}, W.~{Saad}, M.~{Bennis}, and M.~{Debbah}, ``{Integrated Millimeter
  Wave and Sub-6 GHz Wireless Networks: A Roadmap for Joint Mobile Broadband
  and Ultra-Reliable Low-Latency Communications},'' \emph{IEEE Wireless
  Communications}, vol.~26, no.~2, pp. 109--115, April 2019.

\bibitem{carvalho17}
E.~d.~{Carvalho}, E.~{Bjornson}, J.~H. {Sorensen}, P.~{Popovski}, and E.~G.
  {Larsson}, ``{Random Access Protocols for Massive MIMO},'' \emph{IEEE
  Communications Magazine}, vol.~55, no.~5, pp. 216--222, May 2017.

\bibitem{popovski14}
P.~{Popovski}, ``{Ultra-reliable communication in 5G wireless systems},'' in
  \emph{1st International Conference on 5G for Ubiquitous Connectivity}, Nov
  2014, pp. 146--151.

\bibitem{green04}
P.~E. {Green}, ``{Fiber to the home: the next big broadband thing},''
  \emph{IEEE Communications Magazine}, vol.~42, no.~9, pp. 100--106, Sep. 2004.

\bibitem{liu13}
{Cheng Liu}, N.~{Cvijetic}, K.~{Sundaresan}, {Meilong Jiang}, S.~{Rangarajan},
  {Ting Wang}, and {Gee-Kung Chang}, ``{A novel in-building small-cell backhaul
  architecture for cost-efficient multi-operator multi-service coexistence},''
  in \emph{2013 Optical Fiber Communication Conference and Exposition and the
  National Fiber Optic Engineers Conference (OFC/NFOEC)}, March 2013, pp. 1--3.

\bibitem{zhu13}
M.~{Zhu}, L.~{Zhang}, J.~{Wang}, L.~{Cheng}, C.~{Liu}, and G.~{Chang},
  ``{Radio-Over-Fiber Access Architecture for Integrated Broadband Wireless
  Services},'' \emph{Journal of Lightwave Technology}, vol.~31, no.~23, pp.
  3614--3620, Dec 2013.

\bibitem{dat14}
P.~T. {Dat}, A.~{Kanno}, K.~{Inagaki}, and T.~{Kawanishi}, ``{High-Capacity
  Wireless Backhaul Network Using Seamless Convergence of Radio-over-Fiber and
  90-GHz Millimeter-Wave},'' \emph{Journal of Lightwave Technology}, vol.~32,
  no.~20, pp. 3910--3923, Oct 2014.

\bibitem{liu13b}
C.~{Liu}, L.~{Zhang}, M.~{Zhu}, J.~{Wang}, L.~{Cheng}, and G.~{Chang}, ``{A
  Novel Multi-Service Small-Cell Cloud Radio Access Network for Mobile Backhaul
  and Computing Based on Radio-Over-Fiber Technologies},'' \emph{Journal of
  Lightwave Technology}, vol.~31, no.~17, pp. 2869--2875, Sep. 2013.

\bibitem{chang13}
{Gee-Kung Chang}, C.~{Liu}, and {Liang Zhang}, ``{Architecture and applications
  of a versatile small-cell, multi-service cloud radio access network using
  radio-over-fiber technologies},'' in \emph{2013 IEEE International Conference
  on Communications Workshops (ICC)}, June 2013, pp. 879--883.

\bibitem{dat16}
P.~T. {Dat}, A.~{Kanno}, N.~{Yamamoto}, and T.~{Kawanishi}, ``{Simultaneous
  transmission of 4G LTE-A and wideband MMW OFDM signals over fiber links},''
  in \emph{2016 IEEE International Topical Meeting on Microwave Photonics
  (MWP)}, Oct 2016, pp. 87--90.

\bibitem{kanno12}
A.~{Kanno}, P.~T. {Dat}, T.~{Kuri}, I.~{Hosako}, T.~{Kawanishi}, Y.~{Yoshida},
  Y.~{Yasumura}, and K.~{Kitayama}, ``{Coherent Radio-Over-Fiber and
  Millimeter-Wave Radio Seamless Transmission System for Resilient Access
  Networks},'' \emph{IEEE Photonics Journal}, vol.~4, no.~6, pp. 2196--2204,
  Dec 2012.

\bibitem{dat16b}
P.~T. {Dat}, A.~{Kanno}, N.~{Yamamoto}, and T.~{Kawanishi}, ``{Efficient mobile
  fronthaul for simultaneous transmission of 4G and future mobile signals},''
  in \emph{2016 21st OptoElectronics and Communications Conference (OECC) held
  jointly with 2016 International Conference on Photonics in Switching (PS)},
  July 2016, pp. 1--3.

\bibitem{huang17b}
C.~{Huang} and W.~{Lin}, ``{A radio transceiver architecture for coexistence of
  4G-LTE and 5G systems used in mobile devices},'' in \emph{2017 IEEE MTT-S
  International Microwave Symposium (IMS)}, June 2017, pp. 2056--2058.

\bibitem{levanen14}
T.~{Levanen}, J.~{Talvitie}, J.~{Pirskanen}, and M.~{Valkama}, ``{New
  Spectrally and Energy Efficient Flexible TDD Based Air Interface for 5G Small
  Cells},'' in \emph{2014 IEEE 79th Vehicular Technology Conference (VTC
  Spring)}, May 2014, pp. 1--7.

\bibitem{ren18}
Z.~{Ren}, S.~{Wu}, and A.~{Zhao}, ``{Coexist Design of Sub-6GHz and
  Millimeter-Wave Antennas for 5G Mobile Terminals},'' in \emph{2018
  International Symposium on Antennas and Propagation (ISAP)}, Oct 2018, pp.
  1--2.

\bibitem{zhao18}
A.~{Zhao} and F.~{Ai}, ``{Dual-band 5G millimeter-wave MIMO antenna array for
  mobile phone application},'' in \emph{12th European Conference on Antennas
  and Propagation (EuCAP 2018)}, April 2018, pp. 1--5.

\bibitem{li16b}
M.~{Li}, Y.~{Ban}, Z.~{Xu}, G.~{Wu}, C.~{Sim}, K.~{Kang}, and Z.~{Yu},
  ``{Eight-Port Orthogonally Dual-Polarized Antenna Array for 5G Smartphone
  Applications},'' \emph{IEEE Transactions on Antennas and Propagation},
  vol.~64, no.~9, pp. 3820--3830, Sep. 2016.

\bibitem{ban16}
Y.~{Ban}, C.~{Li}, C.~{Sim}, G.~{Wu}, and K.~{Wong}, ``{4G/5G Multiple Antennas
  for Future Multi-Mode Smartphone Applications},'' \emph{IEEE Access}, vol.~4,
  pp. 2981--2988, 2016.

\bibitem{zheng18}
S.~Y. {Zheng}, ``{A Dual-Band Antenna accross Microwave and Millimeter-wave
  Frequency Bands},'' in \emph{2018 International Applied Computational
  Electromagnetics Society Symposium - China (ACES)}, July 2018, pp. 1--2.

\bibitem{yang06}
{Songnan Yang}, S.~H. {Suleiman}, and A.~E. {Fathy}, ``{Ku-band slot array
  antennas for low profile mobile DBS applications: printed vs. machined},'' in
  \emph{2006 IEEE Antennas and Propagation Society International Symposium},
  July 2006, pp. 3137--3140.

\bibitem{chew82}
{Weng Chew}, ``{A broad-band annular-ring microstrip antenna},'' \emph{IEEE
  Transactions on Antennas and Propagation}, vol.~30, no.~5, pp. 918--922, Sep.
  1982.

\bibitem{tsai10}
M.~{Tsai}, P.~{You}, T.~{Tsai}, K.~{Huang}, and T.~{Huang}, ``{Design of
  wide-IF-bandwidth down-conversion mixer for 60-GHz WPAN / 3–10 GHz UWB
  group-1 coexistence system application},'' in \emph{2010 International
  Symposium on Next Generation Electronics}, Nov 2010, pp. 116--119.

\bibitem{gilbert68}
B.~{Gilbert}, ``{A precise four-quadrant multiplier with subnanosecond
  response},'' \emph{IEEE Journal of Solid-State Circuits}, vol.~3, no.~4, pp.
  365--373, Dec 1968.

\bibitem{ye18}
X.~F. {Ye}, S.~Y. {Zheng}, Y.~M. {Pan}, D.~{Ho}, and Y.~{Long}, ``{A New Class
  of Components for Simultaneous Power Splitting Over Microwave and
  Millimeter-Wave Frequency Bands},'' \emph{IEEE Access}, vol.~6, pp. 146--158,
  2018.

\bibitem{woo06}
{Duk-Jae Woo}, {Taek-Kyung Lee}, {Jae-Wook Lee}, {Cheol-Sig Pyo}, and {Won-Kyu
  Choi}, ``{Novel U-slot and V-slot DGSs for bandstop filter with improved Q
  factor},'' \emph{IEEE Transactions on Microwave Theory and Techniques},
  vol.~54, no.~6, pp. 2840--2847, June 2006.

\bibitem{garg13}
R.~Garg, I.~Bahl, and M.~Bozzi, \emph{{Microstrip lines and slotlines}}.\hskip
  1em plus 0.5em minus 0.4em\relax Artech house, 2013.

\bibitem{wells10}
J.~Wells, \emph{{Multi-gigabit microwave and millimeter-wave wireless
  communications}}.\hskip 1em plus 0.5em minus 0.4em\relax Artech House, 2010.

\bibitem{semiari16}
O.~{Semiari}, W.~{Saad}, and M.~{Bennis}, ``{Context-aware scheduling of joint
  millimeter wave and microwave resources for dual-mode base stations},'' in
  \emph{2016 IEEE International Conference on Communications (ICC)}, May 2016,
  pp. 1--6.

\bibitem{semiari17}
------, ``{Joint Millimeter Wave and Microwave Resources Allocation in Cellular
  Networks With Dual-Mode Base Stations},'' \emph{IEEE Transactions on Wireless
  Communications}, vol.~16, no.~7, pp. 4802--4816, July 2017.

\bibitem{ghosh14}
A.~{Ghosh}, T.~A. {Thomas}, M.~C. {Cudak}, R.~{Ratasuk}, P.~{Moorut}, F.~W.
  {Vook}, T.~S. {Rappaport}, G.~R. {MacCartney}, S.~{Sun}, and S.~{Nie},
  ``{Millimeter-Wave Enhanced Local Area Systems: A High-Data-Rate Approach for
  Future Wireless Networks},'' \emph{IEEE Journal on Selected Areas in
  Communications}, vol.~32, no.~6, pp. 1152--1163, June 2014.

\bibitem{ghosh09}
A.~{Ghosh} and R.~{Ratasuk}, ``{Multi-Antenna Systems for LTE eNodeB},'' in
  \emph{2009 IEEE 70th Vehicular Technology Conference Fall}, Sep. 2009, pp.
  1--4.

\bibitem{roth92}
A.~E. Roth and M.~Sotomayor, ``{Two-sided matching},'' \emph{Handbook of game
  theory with economic applications}, vol.~1, pp. 485--541, 1992.

\bibitem{jorswieck11}
E.~A. {Jorswieck}, ``{Stable matchings for resource allocation in wireless
  networks},'' in \emph{2011 17th International Conference on Digital Signal
  Processing (DSP)}, July 2011, pp. 1--8.

\bibitem{semiari14}
O.~{Semiari}, W.~{Saad}, S.~{Valentin}, M.~{Bennis}, and B.~{Maham},
  ``{Matching theory for priority-based cell association in the downlink of
  wireless small cell networks},'' in \emph{2014 IEEE International Conference
  on Acoustics, Speech and Signal Processing (ICASSP)}, May 2014, pp. 444--448.

\bibitem{semiari15}
O.~{Semiari}, W.~{Saad}, S.~{Valentin}, M.~{Bennis}, and H.~V. {Poor},
  ``{Context-Aware Small Cell Networks: How Social Metrics Improve Wireless
  Resource Allocation},'' \emph{IEEE Transactions on Wireless Communications},
  vol.~14, no.~11, pp. 5927--5940, Nov 2015.

\bibitem{gu15}
Y.~{Gu}, W.~{Saad}, M.~{Bennis}, M.~{Debbah}, and Z.~{Han}, ``{Matching theory
  for future wireless networks: fundamentals and applications},'' \emph{IEEE
  Communications Magazine}, vol.~53, no.~5, pp. 52--59, May 2015.

\bibitem{watkins92}
C.~J. Watkins and P.~Dayan, ``{Q-learning},'' \emph{Machine learning}, vol.~8,
  no. 3-4, pp. 279--292, 1992.

\bibitem{simsek12}
M.~{Simsek}, M.~{Bennis}, and A.~{Czylwik}, ``{Dynamic Inter-Cell Interference
  Coordination in HetNets: A reinforcement learning approach},'' in \emph{2012
  IEEE Global Communications Conference (GLOBECOM)}, Dec 2012, pp. 5446--5450.

\bibitem{sutton18}
R.~S. Sutton and A.~G. Barto, \emph{{Reinforcement learning: An
  introduction}}.\hskip 1em plus 0.5em minus 0.4em\relax MIT press, 2018.

\bibitem{dean05}
B.~C. Dean, ``{Approximation algorithms for stochastic scheduling problems},''
  Ph.D. dissertation, Massachusetts Institute of Technology, 2005.

\bibitem{chergui19}
H.~{Chergui}, K.~{Tourki}, R.~{Lguensat}, M.~{Benjillali}, C.~{Verikoukis}, and
  M.~{Debbah}, ``{Classification Algorithms for Semi-Blind Uplink/Downlink
  Decoupling in Sub-6 GHz/mmWave 5G Networks},'' in \emph{2019 15th
  International Wireless Communications Mobile Computing Conference (IWCMC)},
  June 2019, pp. 2031--2035.

\bibitem{jaeckel14}
S.~{Jaeckel}, L.~{Raschkowski}, K.~{Börner}, and L.~{Thiele}, ``{QuaDRiGa: A
  3-D Multi-Cell Channel Model With Time Evolution for Enabling Virtual Field
  Trials},'' \emph{IEEE Transactions on Antennas and Propagation}, vol.~62,
  no.~6, pp. 3242--3256, June 2014.

\bibitem{atzeni18}
I.~{Atzeni}, J.~{Arnau}, and M.~{Kountouris}, ``{Downlink Cellular Network
  Analysis With LOS/NLOS Propagation and Elevated Base Stations},'' \emph{IEEE
  Transactions on Wireless Communications}, vol.~17, no.~1, pp. 142--156, Jan
  2018.

\bibitem{sesia11}
S.~Sesia, I.~Toufik, and M.~Baker, \emph{{LTE-the UMTS long term evolution:
  from theory to practice}}.\hskip 1em plus 0.5em minus 0.4em\relax John Wiley
  \& Sons, 2011.

\bibitem{tipping99}
M.~E. Tipping and C.~M. Bishop, ``{Probabilistic principal component
  analysis},'' \emph{Journal of the Royal Statistical Society: Series B
  (Statistical Methodology)}, vol.~61, no.~3, pp. 611--622, 1999.

\bibitem{cortes95}
C.~Cortes and V.~Vapnik, ``{Support-vector networks},'' \emph{Machine
  learning}, vol.~20, no.~3, pp. 273--297, 1995.

\bibitem{park08}
M.~{Park}, C.~{Cordeiro}, E.~{Perahia}, and L.~L. {Yang}, ``{Millimeter-wave
  multi-Gigabit WLAN: Challenges and feasibility},'' in \emph{2008 IEEE 19th
  International Symposium on Personal, Indoor and Mobile Radio Communications},
  Sep. 2008, pp. 1--5.

\bibitem{semiari17b}
O.~{Semiari}, W.~{Saad}, M.~{Bennis}, and M.~{Debbah}, ``{Performance Analysis
  of Integrated Sub-6 GHz-Millimeter Wave Wireless Local Area Networks},'' in
  \emph{GLOBECOM 2017 - 2017 IEEE Global Communications Conference}, Dec 2017,
  pp. 1--7.

\bibitem{nguyen17}
K.~{Nguyen}, M.~G. {Kibria}, K.~{Ishizu}, and F.~{Kojima}, ``{Feasibility Study
  of Providing Backward Compatibility with MPTCP to WiGig/IEEE 802.11ad},'' in
  \emph{2017 IEEE 86th Vehicular Technology Conference (VTC-Fall)}, Sep. 2017,
  pp. 1--5.

\bibitem{ford12}
A.~Ford, C.~Raiciu, M.~Handley, O.~Bonaventure \emph{et~al.}, ``{TCP Extensions
  for Multipath Operation with Multiple Addresses,
  draft-ietf-mptcp-multiaddressed-09},'' \emph{Internetdraft, IETF (March
  2012)}, 2012.

\bibitem{paasch12}
C.~Paasch, G.~Detal, F.~Duchene, C.~Raiciu, and O.~Bonaventure, ``{Exploring
  mobile/WiFi handover with multipath TCP},'' in \emph{Proceedings of the 2012
  ACM SIGCOMM workshop on Cellular networks: operations, challenges, and future
  design}.\hskip 1em plus 0.5em minus 0.4em\relax ACM, 2012, pp. 31--36.

\bibitem{chen13}
Y.-C. Chen, Y.-s. Lim, R.~J. Gibbens, E.~M. Nahum, R.~Khalili, and D.~Towsley,
  ``{A measurement-based study of multipath tcp performance over wireless
  networks},'' in \emph{Proceedings of the 2013 conference on Internet
  measurement conference}.\hskip 1em plus 0.5em minus 0.4em\relax ACM, 2013,
  pp. 455--468.

\bibitem{lim14}
Y.~{Lim}, Y.~{Chen}, E.~M. {Nahum}, D.~{Towsley}, and K.~{Lee}, ``{Cross-layer
  path management in multi-path transport protocol for mobile devices},'' in
  \emph{IEEE INFOCOM 2014 - IEEE Conference on Computer Communications}, April
  2014, pp. 1815--1823.

\bibitem{nguyen14}
K.~{Nguyen}, Y.~{Ji}, and S.~{Yamada}, ``{A cross-layer approach for improving
  WiFi performance},'' in \emph{2014 International Wireless Communications and
  Mobile Computing Conference (IWCMC)}, Aug 2014, pp. 458--463.

\bibitem{apostolidis13}
A.~Apostolidis, L.~Campoy, K.~Chatzikokolakis, K.~Friederichs, T.~Irnich,
  K.~Koufos, J.~Kronander, J.~Luo, E.~Mohyeldin, P.~Olmos \emph{et~al.},
  ``{Intermediate description of the spectrum needs and usage principles},''
  \emph{METIS Deliverable D}, vol.~5, p.~1, 2013.

\bibitem{liolis13}
K.~{Liolis}, G.~{Schlueter}, J.~{Krause}, F.~{Zimmer}, L.~{Combelles},
  J.~{Grotz}, S.~{Chatzinotas}, B.~{Evans}, A.~{Guidotti}, D.~{Tarchi}, and
  A.~{Vanelli-Coralli}, ``{Cognitive radio scenarios for satellite
  communications: The CoRaSat approach},'' in \emph{2013 Future Network Mobile
  Summit}, July 2013, pp. 1--10.

\bibitem{ippolito17}
L.~J. Ippolito, \emph{{Satellite communications systems engineering}}.\hskip
  1em plus 0.5em minus 0.4em\relax Wiley Online Library, 2017.

\bibitem{guidolin15}
F.~{Guidolin}, M.~{Nekovee}, L.~{Badia}, and M.~{Zorzi}, ``{A cooperative
  scheduling algorithm for the coexistence of fixed satellite services and 5G
  cellular network},'' in \emph{2015 IEEE International Conference on
  Communications (ICC)}, June 2015, pp. 1322--1327.

\bibitem{guidolin15b}
------, ``{A study on the coexistence of fixed satellite service and cellular
  networks in a mmWave scenario},'' in \emph{2015 IEEE International Conference
  on Communications (ICC)}, June 2015, pp. 2444--2449.

\bibitem{guidolin15c}
F.~{Guidolin} and M.~{Nekovee}, ``Investigating spectrum sharing between 5g
  millimeter wave networks and fixed satellite systems,'' in \emph{2015 IEEE
  Globecom Workshops (GC Wkshps)}, Dec 2015, pp. 1--7.

\bibitem{bai14}
T.~{Bai}, V.~{Desai}, and R.~W. {Heath}, ``{Millimeter wave cellular channel
  models for system evaluation},'' in \emph{2014 International Conference on
  Computing, Networking and Communications (ICNC)}, Feb 2014, pp. 178--182.

\bibitem{hur15}
S.~{Hur}, Y.~{Cho}, {Taehwan Kim}, J.~{Park}, A.~F. {Molisch}, K.~{Haneda}, and
  M.~{Peter}, ``{Wideband spatial channel model in an urban cellular
  environments at 28 GHz},'' in \emph{2015 9th European Conference on Antennas
  and Propagation (EuCAP)}, April 2015, pp. 1--5.

\bibitem{monderer96}
D.~Monderer and L.~S. Shapley, ``{Potential games},'' \emph{Games and economic
  behavior}, vol.~14, no.~1, pp. 124--143, 1996.

\bibitem{nie06}
N.~Nie and C.~Comaniciu, ``{Adaptive channel allocation spectrum etiquette for
  cognitive radio networks},'' \emph{Mobile networks and applications},
  vol.~11, no.~6, pp. 779--797, 2006.

\bibitem{buzzi12}
S.~{Buzzi}, G.~{Colavolpe}, D.~{Saturnino}, and A.~{Zappone}, ``{Potential
  Games for Energy-Efficient Power Control and Subcarrier Allocation in Uplink
  Multicell OFDMA Systems},'' \emph{IEEE Journal of Selected Topics in Signal
  Processing}, vol.~6, no.~2, pp. 89--103, April 2012.

\bibitem{chen13b}
F.~{Chen} and J.~{Kao}, ``{Game-Based Broadcast over Reliable and Unreliable
  Wireless Links in Wireless Multihop Networks},'' \emph{IEEE Transactions on
  Mobile Computing}, vol.~12, no.~8, pp. 1613--1624, Aug 2013.

\bibitem{mogensen07}
P.~{Mogensen}, W.~{Na}, I.~Z. {Kovacs}, F.~{Frederiksen}, A.~{Pokhariyal},
  K.~I. {Pedersen}, T.~{Kolding}, K.~{Hugl}, and M.~{Kuusela}, ``{LTE Capacity
  Compared to the Shannon Bound},'' in \emph{2007 IEEE 65th Vehicular
  Technology Conference - VTC2007-Spring}, April 2007, pp. 1234--1238.

\bibitem{kim17}
S.~{Kim}, E.~{Visotsky}, P.~{Moorut}, K.~{Bechta}, A.~{Ghosh}, and
  C.~{Dietrich}, ``{Coexistence of 5G With the Incumbents in the 28 and 70 GHz
  Bands},'' \emph{IEEE Journal on Selected Areas in Communications}, vol.~35,
  no.~6, pp. 1254--1268, June 2017.

\bibitem{chiu13}
S.~N. Chiu, D.~Stoyan, W.~S. Kendall, and J.~Mecke, \emph{{Stochastic geometry
  and its applications}}.\hskip 1em plus 0.5em minus 0.4em\relax John Wiley \&
  Sons, 2013.

\bibitem{anderson03}
H.~R. Anderson, \emph{{Fixed broadband wireless system design}}.\hskip 1em plus
  0.5em minus 0.4em\relax John Wiley \& Sons, 2003.

\bibitem{lin18}
Z.~{Lin}, M.~{Lin}, J.~{Wang}, Y.~{Huang}, and W.~{Zhu}, ``{Robust Secure
  Beamforming for 5G Cellular Networks Coexisting With Satellite Networks},''
  \emph{IEEE Journal on Selected Areas in Communications}, vol.~36, no.~4, pp.
  932--945, April 2018.

\bibitem{sharma13}
S.~K. {Sharma}, S.~{Chatzinotas}, and B.~{Ottersten}, ``{Transmit beamforming
  for spectral coexistence of satellite and terrestrial networks},'' in
  \emph{8th International Conference on Cognitive Radio Oriented Wireless
  Networks}, July 2013, pp. 275--281.

\bibitem{vassaki13}
S.~{Vassaki}, M.~I. {Poulakis}, A.~D. {Panagopoulos}, and P.~{Constantinou},
  ``{Power Allocation in Cognitive Satellite Terrestrial Networks with QoS
  Constraints},'' \emph{IEEE Communications Letters}, vol.~17, no.~7, pp.
  1344--1347, July 2013.

\bibitem{an16}
K.~{An}, M.~{Lin}, T.~{Liang}, J.~{Ouyang}, and W.~{Zhu}, ``{On the ergodic
  capacity of multiple antenna cognitive satellite terrestrial networks},'' in
  \emph{2016 IEEE International Conference on Communications (ICC)}, May 2016,
  pp. 1--5.

\bibitem{an16b}
K.~{An}, M.~{Lin}, W.~{Zhu}, Y.~{Huang}, and G.~{Zheng}, ``{Outage Performance
  of Cognitive Hybrid Satellite–Terrestrial Networks With Interference
  Constraint},'' \emph{IEEE Transactions on Vehicular Technology}, vol.~65,
  no.~11, pp. 9397--9404, Nov 2016.

\bibitem{an16c}
K.~{An}, M.~{Lin}, J.~{Ouyang}, and W.~{Zhu}, ``{Secure Transmission in
  Cognitive Satellite Terrestrial Networks},'' \emph{IEEE Journal on Selected
  Areas in Communications}, vol.~34, no.~11, pp. 3025--3037, Nov 2016.

\bibitem{an17}
K.~{An}, M.~{Lin}, J.~{Guyang}, T.~{Liang}, J.~{Wang}, and W.~{Zhu}, ``{Outage
  performance for the cognitive broadband satellite system and terrestrial
  cellular network in millimeter wave scenario},'' in \emph{2017 IEEE
  International Conference on Communications (ICC)}, May 2017, pp. 1--6.

\bibitem{rappaport96}
T.~S. Rappaport \emph{et~al.}, \emph{{Wireless communications: principles and
  practice}}.\hskip 1em plus 0.5em minus 0.4em\relax prentice hall PTR New
  Jersey, 1996, vol.~2.

\bibitem{zou14}
Y.~{Zou}, X.~{Li}, and Y.~{Liang}, ``{Secrecy Outage and Diversity Analysis of
  Cognitive Radio Systems},'' \emph{IEEE Journal on Selected Areas in
  Communications}, vol.~32, no.~11, pp. 2222--2236, November 2014.

\bibitem{meng19}
X.~Meng, L.~Zhong, D.~Zhou, and D.~Yang, ``{Co-Channel Coexistence Analysis
  between 5G IoT System and Fixed-Satellite Service at 40 GHz},''
  \emph{Wireless Communications and Mobile Computing}, vol. 2019, 2019.

\bibitem{hong14}
W.~{Hong}, K.~{Baek}, Y.~{Lee}, Y.~{Kim}, and S.~{Ko}, ``{Study and prototyping
  of practically large-scale mmWave antenna systems for 5G cellular devices},''
  \emph{IEEE Communications Magazine}, vol.~52, no.~9, pp. 63--69, Sep. 2014.

\bibitem{han19}
S.~{Han}, Y.~{Huang}, W.~{Meng}, C.~{Li}, N.~{Xu}, and D.~{Chen}, ``{Optimal
  Power Allocation for SCMA Downlink Systems Based on Maximum Capacity},''
  \emph{IEEE Transactions on Communications}, vol.~67, no.~2, pp. 1480--1489,
  Feb 2019.

\bibitem{federal16}
F.~C. Commission \emph{et~al.}, ``{Use of spectrum bands above 24 GHz for
  mobile radio services},'' \emph{Fed Regist}, vol.~81, no. 164, pp.
  58\,270--58\,308, 2016.

\bibitem{hattab18}
G.~{Hattab}, E.~{Visotsky}, M.~{Cudak}, and A.~{Ghosh}, ``{Toward the
  Coexistence of 5G MmWave Networks with Incumbent Systems beyond 70 GHz},''
  \emph{IEEE Wireless Communications}, vol.~25, no.~4, pp. 18--24, AUGUST 2018.

\bibitem{hattab17}
------, ``{Coexistence of 5G mmWave Users with Incumbent Fixed Stations over 70
  and 80 GHz},'' in \emph{2017 IEEE Globecom Workshops (GC Wkshps)}, Dec 2017,
  pp. 1--5.

\bibitem{hattab19}
G.~{Hattab}, E.~{Visotsky}, M.~C. {Cudak}, and A.~{Ghosh}, ``{Uplink
  Interference Mitigation Techniques for Coexistence of 5G Millimeter Wave
  Users With Incumbents at 70 and 80 GHz},'' \emph{IEEE Transactions on
  Wireless Communications}, vol.~18, no.~1, pp. 324--339, Jan 2019.

\bibitem{hattab18b}
G.~{Hattab}, E.~{Visotsky}, M.~{Cudak}, and A.~{Ghosh}, ``{Interference
  Mitigation via Beam Range Biasing for 5G mmWave Coexistence with
  Incumbents},'' in \emph{2018 IEEE 5G World Forum (5GWF)}, July 2018, pp.
  210--214.

\bibitem{kim15}
{Joongheon Kim}, {Liang Xian}, A.~{Maltsev}, R.~{Arefi}, and A.~S. {Sadri},
  ``{Study of coexistence between 5G small-cell systems and systems of the
  fixed service at 39 GHz band},'' in \emph{2015 IEEE MTT-S International
  Microwave Symposium}, May 2015, pp. 1--3.

\bibitem{choi18}
S.~{Choi}, S.~{Park}, and K.~K. {Radio}, ``{Coexistence analysis between
  intelligent transport systems and wireless multi-gigabit service in 60GHz
  bands},'' in \emph{2018 International Conference on Information and
  Communication Technology Convergence (ICTC)}, Oct 2018, pp. 1200--1203.

\bibitem{administrations99}
CEPT, ``{A comparison of the minimum coupling loss method, enhanced minimum
  coupling loss method, and the monte-carlo simulation (ERC Report101)},''
  \emph{ERC}, May 1999.

\bibitem{ding17b}
Z.~{Ding}, P.~{Fan}, and H.~V. {Poor}, ``{On the coexistence of non-orthogonal
  multiple access and millimeter-wave communications},'' in \emph{2017 IEEE
  International Conference on Communications (ICC)}, May 2017, pp. 1--6.

\bibitem{haenggi12}
M.~Haenggi, \emph{{Stochastic geometry for wireless networks}}.\hskip 1em plus
  0.5em minus 0.4em\relax Cambridge University Press, 2012.

\bibitem{rappaport12}
T.~S. {Rappaport}, E.~{Ben-Dor}, J.~N. {Murdock}, and Y.~{Qiao}, ``{38 GHz and
  60 GHz angle-dependent propagation for cellular peer-to-peer wireless
  communications},'' in \emph{2012 IEEE International Conference on
  Communications (ICC)}, June 2012, pp. 4568--4573.

\bibitem{lee16}
G.~{Lee}, Y.~{Sung}, and J.~{Seo}, ``{Randomly-Directional Beamforming in
  Millimeter-Wave Multiuser MISO Downlink},'' \emph{IEEE Transactions on
  Wireless Communications}, vol.~15, no.~2, pp. 1086--1100, Feb 2016.

\bibitem{andrews10}
J.~G. {Andrews}, R.~K. {Ganti}, M.~{Haenggi}, N.~{Jindal}, and S.~{Weber}, ``A
  primer on spatial modeling and analysis in wireless networks,'' \emph{IEEE
  Communications Magazine}, vol.~48, no.~11, pp. 156--163, November 2010.

\bibitem{thornburg16}
A.~{Thornburg}, T.~{Bai}, and R.~W. {Heath}, ``{Performance Analysis of Outdoor
  mmWave Ad Hoc Networks},'' \emph{IEEE Transactions on Signal Processing},
  vol.~64, no.~15, pp. 4065--4079, Aug 2016.

\bibitem{sun15}
Q.~{Sun}, S.~{Han}, C.~{I}, and Z.~{Pan}, ``{On the Ergodic Capacity of MIMO
  NOMA Systems},'' \emph{IEEE Wireless Communications Letters}, vol.~4, no.~4,
  pp. 405--408, Aug 2015.

\bibitem{ding16}
Z.~{Ding}, P.~{Fan}, and H.~V. {Poor}, ``{Impact of User Pairing on 5G
  Nonorthogonal Multiple-Access Downlink Transmissions},'' \emph{IEEE
  Transactions on Vehicular Technology}, vol.~65, no.~8, pp. 6010--6023, Aug
  2016.

\bibitem{lee16b}
G.~{Lee}, Y.~{Sung}, and M.~{Kountouris}, ``{On the Performance of Random
  Beamforming in Sparse Millimeter Wave Channels},'' \emph{IEEE Journal of
  Selected Topics in Signal Processing}, vol.~10, no.~3, pp. 560--575, April
  2016.

\bibitem{gao16}
X.~{Gao}, L.~{Dai}, S.~{Han}, C.~{I}, and R.~W. {Heath}, ``{Energy-Efficient
  Hybrid Analog and Digital Precoding for MmWave MIMO Systems With Large
  Antenna Arrays},'' \emph{IEEE Journal on Selected Areas in Communications},
  vol.~34, no.~4, pp. 998--1009, April 2016.

\bibitem{zhou18}
F.~{Zhou}, Y.~{Wu}, R.~Q. {Hu}, Y.~{Wang}, and K.~K. {Wong},
  ``{Energy-Efficient NOMA Enabled Heterogeneous Cloud Radio Access
  Networks},'' \emph{IEEE Network}, vol.~32, no.~2, pp. 152--160, March 2018.

\bibitem{kusaladharma18}
S.~{Kusaladharma}, W.~P. {Zhu}, and W.~{Ajib}, ``{Downlink NOMA for Stochastic
  Cellular Networks under Millimeter Wave Channels},'' in \emph{2018 IEEE
  Global Communications Conference (GLOBECOM)}, Dec 2018, pp. 1--6.

\bibitem{zhang17b}
Z.~{Zhang}, Z.~{Ma}, Y.~{Xiao}, M.~{Xiao}, G.~K. {Karagiannidis}, and P.~{Fan},
  ``{Non-Orthogonal Multiple Access for Cooperative Multicast Millimeter Wave
  Wireless Networks},'' \emph{IEEE Journal on Selected Areas in
  Communications}, vol.~35, no.~8, pp. 1794--1808, Aug 2017.

\bibitem{lin14b}
X.~{Lin}, R.~{Ratasuk}, A.~{Ghosh}, and J.~G. {Andrews}, ``{Modeling, Analysis,
  and Optimization of Multicast Device-to-Device Transmissions},'' \emph{IEEE
  Transactions on Wireless Communications}, vol.~13, no.~8, pp. 4346--4359, Aug
  2014.

\bibitem{kim16}
J.~{Kim}, S.~W. {Choi}, W.~{Shin}, Y.~{Song}, and Y.~{Kim}, ``{Group
  communication over LTE: a radio access perspective},'' \emph{IEEE
  Communications Magazine}, vol.~54, no.~4, pp. 16--23, April 2016.

\bibitem{fuente16}
A.~{de la Fuente}, R.~P. {Leal}, and A.~G. {Armada}, ``{New Technologies and
  Trends for Next Generation Mobile Broadcasting Services},'' \emph{IEEE
  Communications Magazine}, vol.~54, no.~11, pp. 217--223, November 2016.

\bibitem{condoluci16}
M.~{Condoluci}, G.~{Araniti}, T.~{Mahmoodi}, and M.~{Dohler}, ``{Enabling the
  IoT Machine Age With 5G: Machine-Type Multicast Services for Innovative
  Real-Time Applications},'' \emph{IEEE Access}, vol.~4, pp. 5555--5569, 2016.

\bibitem{araniti17}
G.~{Araniti}, M.~{Condoluci}, P.~{Scopelliti}, A.~{Molinaro}, and A.~{Iera},
  ``{Multicasting over Emerging 5G Networks: Challenges and Perspectives},''
  \emph{IEEE Network}, vol.~31, no.~2, pp. 80--89, March 2017.

\bibitem{morgado17}
A.~J. {Morgado}, K.~M.~S. {Huq}, J.~{Rodriguez}, C.~{Politis}, and
  H.~{Gacanin}, ``{Hybrid Resource Allocation for Millimeter-Wave NOMA},''
  \emph{IEEE Wireless Communications}, vol.~24, no.~5, pp. 23--29, October
  2017.

\bibitem{zhu15}
Y.~Zhu, H.-J.~E. Kwon, H.~Jung, U.~Kumar, and J.-k.~J. Fwu, ``{Non-orthogonal
  multiple access (NOMA) wireless systems and methods},'' Oct.~29 2015, uS
  Patent App. 14/632,291.

\bibitem{nekovee16}
M.~{Nekovee}, Y.~{Qi}, and Y.~{Wang}, ``{Distributed beam scheduling for
  multi-RAT coexistence in mm-wave 5G networks},'' in \emph{2016 IEEE 27th
  Annual International Symposium on Personal, Indoor, and Mobile Radio
  Communications (PIMRC)}, Sep. 2016, pp. 1--6.

\bibitem{nekovee17}
M.~Nekovee, Y.~Qi, and Y.~Wang, ``{Self-organized beam scheduling as an enabler
  for coexistence in 5G unlicensed bands},'' \emph{IEICE Transactions on
  Communications}, vol. 100, no.~8, pp. 1181--1189, 2017.

\bibitem{trevisan11}
L.~Trevisan, ``{Combinatorial optimization: exact and approximate
  algorithms},'' \emph{Standford University}, 2011.

\bibitem{bazaraa13}
M.~S. Bazaraa, H.~D. Sherali, and C.~M. Shetty, \emph{{Nonlinear programming:
  theory and algorithms}}.\hskip 1em plus 0.5em minus 0.4em\relax John Wiley \&
  Sons, 2013.

\bibitem{anjum17}
M.~N. {Anjum} and {Hua Fang}, ``{Coexistence in millimeter-wave WBAN: A game
  theoretic approach},'' in \emph{2017 International Conference on Computing,
  Networking and Communications (ICNC)}, Jan 2017, pp. 571--576.

\bibitem{chahat13}
N.~{Chahat}, G.~{Valerio}, M.~{Zhadobov}, and R.~{Sauleau}, ``{On-Body
  Propagation at 60 GHz},'' \emph{IEEE Transactions on Antennas and
  Propagation}, vol.~61, no.~4, pp. 1876--1888, April 2013.

\bibitem{sarimin14}
N.~{Sarimin} and R.~{Abdaoui}, ``{60 GHz channel modeling scenarios and
  characterization for on-body sensors applications},'' in \emph{2014 IEEE
  MTT-S International Microwave Workshop Series on RF and Wireless Technologies
  for Biomedical and Healthcare Applications (IMWS-Bio2014)}, Dec 2014, pp.
  1--3.

\bibitem{fang16}
G.~Fang, M.~A. Orgun, R.~Shankaran, E.~Dutkiewicz, and G.~Zheng, ``{Truthful
  channel sharing for self coexistence of overlapping medical body area
  networks},'' \emph{PloS one}, vol.~11, no.~2, p. e0148376, 2016.

\bibitem{mackenzie06}
A.~B. MacKenzie and L.~A. DaSilva, ``{Game theory for wireless engineers},''
  \emph{Synthesis Lectures on Communications}, vol.~1, no.~1, pp. 1--86, 2006.

\bibitem{bacci15}
G.~{Bacci}, L.~{Sanguinetti}, and M.~{Luise}, ``{Understanding Game Theory via
  Wireless Power Control [Lecture Notes]},'' \emph{IEEE Signal Processing
  Magazine}, vol.~32, no.~4, pp. 132--137, July 2015.

\bibitem{gupta16}
A.~K. {Gupta}, J.~G. {Andrews}, and R.~W. {Heath}, ``{Can operators simply
  share millimeter wave spectrum licenses?}'' in \emph{2016 Information Theory
  and Applications Workshop (ITA)}, Jan 2016, pp. 1--7.

\bibitem{deng18}
N.~{Deng}, M.~{Haenggi}, and Y.~{Sun}, ``{Millimeter-Wave Device-to-Device
  Networks With Heterogeneous Antenna Arrays},'' \emph{IEEE Transactions on
  Communications}, vol.~66, no.~9, pp. 4271--4285, Sep. 2018.

\bibitem{haenggi14}
M.~{Haenggi}, ``{The Mean Interference-to-Signal Ratio and Its Key Role in
  Cellular and Amorphous Networks},'' \emph{IEEE Wireless Communications
  Letters}, vol.~3, no.~6, pp. 597--600, Dec 2014.

\bibitem{balanis16}
C.~A. Balanis, \emph{{Antenna theory: analysis and design}}.\hskip 1em plus
  0.5em minus 0.4em\relax John wiley \& sons, 2016.

\bibitem{singh15}
S.~{Singh}, M.~N. {Kulkarni}, A.~{Ghosh}, and J.~G. {Andrews}, ``{Tractable
  Model for Rate in Self-Backhauled Millimeter Wave Cellular Networks},''
  \emph{IEEE Journal on Selected Areas in Communications}, vol.~33, no.~10, pp.
  2196--2211, Oct 2015.

\bibitem{yu17}
X.~{Yu}, J.~{Zhang}, M.~{Haenggi}, and K.~B. {Letaief}, ``{Coverage Analysis
  for Millimeter Wave Networks: The Impact of Directional Antenna Arrays},''
  \emph{IEEE Journal on Selected Areas in Communications}, vol.~35, no.~7, pp.
  1498--1512, July 2017.

\bibitem{mishra19}
K.~V. {Mishra}, M.~R. {Bhavani Shankar}, V.~{Koivunen}, B.~{Ottersten}, and
  S.~A. {Vorobyov}, ``{Toward Millimeter-Wave Joint Radar Communications: A
  signal processing perspective},'' \emph{IEEE Signal Processing Magazine},
  vol.~36, no.~5, pp. 100--114, Sep. 2019.

\bibitem{cohen18}
D.~{Cohen}, K.~V. {Mishra}, and Y.~C. {Eldar}, ``{Spectrum Sharing Radar:
  Coexistence via Xampling},'' \emph{IEEE Transactions on Aerospace and
  Electronic Systems}, vol.~54, no.~3, pp. 1279--1296, June 2018.

\bibitem{shi04}
{Tingting Shi}, {Shidong Zhou}, and {Yan Yao}, ``{Capacity of single carrier
  systems with frequency-domain equalization},'' in \emph{Proceedings of the
  IEEE 6th Circuits and Systems Symposium on Emerging Technologies: Frontiers
  of Mobile and Wireless Communication (IEEE Cat. No.04EX710)}, vol.~2, May
  2004, pp. 429--432 Vol.2.

\bibitem{takizawa12}
K.~{Takizawa}, M.~{Kyrö}, K.~{Haneda}, H.~{Hagiwara}, and P.~{Vainikainen},
  ``{Performance evaluation of 60 GHz radio systems in hospital
  environments},'' in \emph{2012 IEEE International Conference on
  Communications (ICC)}, June 2012, pp. 3219--3295.

\bibitem{liu13c}
W.~{Liu}, F.~{Yeh}, T.~{Wei}, C.~{Chan}, and S.~{Jou}, ``{A Digital Golay-MPIC
  Time Domain Equalizer for SC/OFDM Dual-Modes at 60 GHz Band},'' \emph{IEEE
  Transactions on Circuits and Systems I: Regular Papers}, vol.~60, no.~10, pp.
  2730--2739, Oct 2013.

\bibitem{mahal17}
J.~A. {Mahal}, A.~{Khawar}, A.~{Abdelhadi}, and T.~C. {Clancy}, ``{Spectral
  Coexistence of MIMO Radar and MIMO Cellular System},'' \emph{IEEE
  Transactions on Aerospace and Electronic Systems}, vol.~53, no.~2, pp.
  655--668, April 2017.

\bibitem{cui18}
Y.~{Cui}, V.~{Koivunen}, and X.~{Jing}, ``{Interference Alignment Based
  Spectrum Sharing for MIMO Radar and Communication Systems},'' in \emph{2018
  IEEE 19th International Workshop on Signal Processing Advances in Wireless
  Communications (SPAWC)}, June 2018, pp. 1--5.

\bibitem{geng18}
Z.~Geng, R.~Xu, H.~Deng, and B.~Himed, ``{Fusion of radar sensing and wireless
  communications by embedding communication signals into the radar transmit
  waveform},'' \emph{IET Radar, Sonar \& Navigation}, vol.~12, no.~6, pp.
  632--640, 2018.

\bibitem{ayyar19}
A.~{Ayyar} and K.~V. {Mishra}, ``{Robust Communications-Centric Coexistence for
  Turbo-Coded OFDM with Non-Traditional Radar Interference Models},'' in
  \emph{2019 IEEE Radar Conference (RadarConf)}, April 2019, pp. 1--6.

\bibitem{olmos08}
J.~J.~V. {Olmos}, T.~{Kuri}, T.~{Sono}, K.~{Tamura}, H.~{Toda}, and
  K.~{Kitayama}, ``{Reconfigurable 2.5-Gb/s Baseband and 60-GHz (155-Mb/s)
  Millimeter-Waveband Radio-Over-Fiber (Interleaving) Access Network},''
  \emph{Journal of Lightwave Technology}, vol.~26, no.~15, pp. 2506--2512, Aug
  2008.

\bibitem{won10}
Y.~{Won}, H.~{Kim}, Y.~{Son}, and S.~{Han}, ``{Full Colorless WDM-Radio Over
  Fiber Access Network Supporting Simultaneous Transmission of Millimeter-Wave
  Band and Baseband Gigabit Signals by Sideband Routing},'' \emph{Journal of
  Lightwave Technology}, vol.~28, no.~16, pp. 2213--2218, Aug 2010.

\bibitem{kuri07}
T.~{Kuri}, H.~{Toda}, and K.~{Kitayama}, ``{Novel Demultiplexer for Dense
  Wavelength-Division-Mutliplexed Millimeter-Wave-Band Radio-Over-Fiber Systems
  With Optical Frequency Interleaving Technique},'' \emph{IEEE Photonics
  Technology Letters}, vol.~19, no.~24, pp. 2018--2020, Dec 2007.

\bibitem{yadav18}
A.~{Yadav} and O.~A. {Dobre}, ``{All Technologies Work Together for Good: A
  Glance at Future Mobile Networks},'' \emph{IEEE Wireless Communications},
  vol.~25, no.~4, pp. 10--16, AUGUST 2018.

\bibitem{bogale16}
T.~E. {Bogale} and L.~B. {Le}, ``{Massive MIMO and mmWave for 5G Wireless
  HetNet: Potential Benefits and Challenges},'' \emph{IEEE Vehicular Technology
  Magazine}, vol.~11, no.~1, pp. 64--75, March 2016.

\bibitem{feng17}
W.~{Feng}, Y.~{Wang}, D.~{Lin}, N.~{Ge}, J.~{Lu}, and S.~{Li}, ``{When mmWave
  Communications Meet Network Densification: A Scalable Interference
  Coordination Perspective},'' \emph{IEEE Journal on Selected Areas in
  Communications}, vol.~35, no.~7, pp. 1459--1471, July 2017.

\bibitem{administrations09}
CEPT, ``{ECC report 113: Compatibility studies around 63GHz between intelligent
  transport system (ITS) and other systems},'' \emph{ECC}, May 2009.

\end{thebibliography}

%





\end{document}